\documentclass[a4paper,11pt]{article}
\usepackage{jheppub} 
\makeatletter
\def\@fpheader{}
\makeatother
\usepackage{microtype}
\usepackage{units}
\usepackage{xcolor}
\usepackage{array}
\usepackage{graphicx}
\usepackage{subcaption}
\usepackage{longtable}
\usepackage{multicol}
\usepackage{comment}
\usepackage{pdfpages}
\usepackage{graphicx}
\usepackage[normalem]{ulem}

\usepackage{breqn}
\usepackage{amsmath}
\usepackage{multirow} 

\usepackage{tikz}
\usetikzlibrary{arrows.meta,positioning,calc,decorations.markings}

\usetikzlibrary{decorations.pathmorphing}

\usepackage{tabularx}
\usepackage{ragged2e}

\usepackage{hyperref}

\newcolumntype{L}{>{$}l<{$}} 
\newcolumntype{C}{>{$}c<{$}}
\newcolumntype{P}{>{$}p<{$}}

\usepackage{orcidlink}

\allowdisplaybreaks

\preprint{CETUP 2025-006}
\title{Large Neutrino-Dark Matter Interactions: 
From Effective Field Theory to Ultraviolet Completions}

\author[]{K.~S.~Babu$^1$\orcidlink
{0000-0001-6147-5155},
}
\affiliation[1]{Department of Physics, Oklahoma State University, Stillwater, OK 74078, USA}
\emailAdd{babu@okstate.edu}

\author[]{P.~S.~Bhupal~Dev$^2$\orcidlink{0000-0003-4655-2866},
}
\affiliation[2]{Department of Physics and McDonnell Center for the Space Sciences, \\
Washington University, St.~Louis, MO 63130, USA}
\emailAdd{bdev@wustl.edu}

\author[]{Anil Thapa$^3$\orcidlink{0000-0003-4471-2336}
}
\affiliation[3]{Physics Department, Colorado State University, Fort Collins, CO 80523, USA}
\emailAdd{a.thapa@colostate.edu}

\abstract{ We develop a general effective field theory (EFT) framework for neutrino-dark matter (DM) interactions,   and apply it to systematically find all possible gauge-invariant ultraviolet (UV) completions at a given EFT operator dimension. Our goal here is to find simple UV-complete models that can realize potentially  large neutrino-DM interactions, while   being consistent with all existing theoretical and experimental constraints. We first construct the leading non-derivative operator basis for neutrino-DM scattering in a low-energy effective theory with neutrinos and DM (DM-LEFT), together with its gauge-invariant embedding in the Standard Model EFT (DM-SMEFT). We then construct all renormalizable tree-level UV completions that generate the relevant DM-SMEFT operators up to dimension-8 using a topology-based classification. Using this framework, we present minimal UV-complete models for different DM types that can yield effective neutrino-DM couplings up to several orders of magnitude larger than the Fermi coupling, while satisfying all constraints, most notably from neutrino mass and from the charged-lepton sector. This includes a pseudo-Dirac fermion DM realization in the scotogenic neutrino mass model and models of Majorana DM inspired by type-II and inverse seesaw-based neutrino mass models. Phenomenological implications for DM thermal relic abundance and direct detection prospects, as well as various cosmological and laboratory constraints on the model parameter space, are also analyzed. 
}

\keywords{Neutrino Interactions, Models for Dark Matter, Effective Field Theories}

\begin{document}

\maketitle

\section{Introduction}
The observation of neutrino oscillations~\cite{Fukuda:1998mi,Ahmad:2002jz,Eguchi:2002dm} provides the first, and so far only, concrete laboratory evidence for the existence of new physics beyond the Standard Model (SM); however, the exact mechanism for neutrino mass generation remains unknown. On the other hand, while the existence of dark matter (DM) in the Universe has been confirmed by numerous astrophysical and cosmological observations~\cite{Bertone:2004pz,Clowe:2006eq,Aghanim:2018eyx}, the particle properties of DM and its non-gravitational interactions with the SM sector are yet to be determined.  
Here we entertain an intriguing possibility that the physics of neutrino mass and of DM are intimately connected. In particular, we would like to address the following question: {\it Can neutrinos have potentially large non-standard interactions with DM, or in general, with dark sector, which might also play a key role in neutrino mass generation, while being consistent with all theoretical and current experimental  constraints?} 

On the experimental/observational front, naively it might  seem like a difficult task to put constraints on the interaction between two `invisible' particles. However, it turns out that there are important astrophysical and cosmological consequences of large neutrino-DM interactions, in particular, for `light' thermal DM with mass roughly below ${\cal O}$(MeV) scale. The most stringent cosmological bounds come from the requirement of neutrino decoupling before the Big Bang Nucleosynthesis (BBN) epoch~\cite{Serpico:2004nm,Mangano:2005cc,Sabti:2019mhn,Giovanetti:2024orj}, as well as from changes in the angular and matter power spectra in Cosmic Microwave Background (CMB) ~\cite{Wilkinson:2014ksa,Escudero:2015yka,DiValentino:2017oaw,Brax:2023tvn} and structure formation  data~\cite{Boehm:2000gq,Mangano:2006mp,Hooper:2021rjc,Akita:2023yga,Heston:2024ljf}. In fact, neutrino-dark sector interactions could potentially resolve some tensions in cosmological datasets~\cite{Farzan:2015pca,Escudero:2022gez,Benso:2024qrg, Zu:2025lrk, Trojanowski:2025oro, Das:2025asx}.  Beyond MeV scale,  the most stringent limits on neutrino-DM (or dark mediator) interactions come from low-energy laboratory experiments, mainly from searches for rare meson decays and $Z$ invisible decay~\cite{Berryman:2018ogk,Brdar:2020nbj,Dev:2024ygx,Dev:2025tdv, Foroughi-Abari:2025mhj,
Okawa:2025sak}, that generally preclude large neutrino-dark sector interactions below ${\cal O}$(GeV) scale.  

On the theoretical front, we can parametrize the neutrino-DM interactions in a minimal, model-independent way using a low-energy effective field theory (LEFT) language, where we only keep the SM neutrinos and the DM fields, and integrate out all other massive SM fields, as well as any other additional heavy fields that might be present in the full theory. However, since neutrinos are part of the same $SU(2)_L$-doublet as the SM charged leptons, it is rather nontrivial to realize large neutrino-DM interactions in a self-consistent, gauge-invariant ultraviolet (UV)-completion of the LEFT operators without running into difficulty with the stringent constraints on DM interactions with the SM charged-lepton sector at the tree-level, and with the SM quark sector at either tree- or loop-level. This is the dark-sector analog of asking whether it is possible to get large neutrino non-standard interactions with matter in a UV-complete framework,  while being consistent with the charged-lepton sector constraints~\cite{Gavela:2008ra,Farzan:2017xzy,Proceedings:2019qno,Babu:2019mfe}. The goal of this paper is to go beyond the simplified model framework for neutrino-DM interactions~\cite{Olivares-DelCampo:2017feq,Blennow:2019fhy, Dev:2025tdv} and to set up an EFT-based general framework to systematically identify all possible UV-completions that can realize potentially large neutrino-DM coupling. For previous EFT analyses involving (sterile) neutrinos and DM, see e.g. Refs.~\cite{Duch:2014xda,GonzalezMacias:2015rxl, Gonzalez-Macias:2016vxy, Criado:2021trs,Coito:2022kif, Liang:2023yta,Borah:2025fkd}. 

In this work, we first construct all leading-order EFT operators in a low-energy EFT description while keeping the DM field as the only non-SM light degree of freedom. We then list the corresponding leading-order gauge-invariant operators in the SM EFT language but including the DM field. We then systematically analyze all possible topologies for UV completion of these EFT operators. Finally, we construct a few simple UV-complete models realizing these operators, and explicitly show that it is indeed possible to achieve relatively large neutrino-DM interaction, while preventing correspondingly large charged lepton-DM interactions at the leading order and large loop-level contributions to neutrino mass. Interestingly, we find that in a UV-complete model, the couplings of the DM to SM charged leptons and quarks are unavoidable and necessarily induced at one-loop level, if not already at tree-level. This leads to interesting direct detection constraints, even for purely ``neutrinophilic" DM models. We discuss the phenomenology of these models and future experimental prospects. In the UV-complete models for fermionic DM   considered here, we find that an effective four-fermion neutrino-DM coupling as large as ${\cal O}(10^5)\times G_F$ is possible, where $G_F$ is the Fermi constant quantifying the strength of the SM weak interaction.

The plan of the paper is as follows: in Section~\ref{sec:2}, we make a comprehensive list of the DM-LEFT and DM-SMEFT operators for different DM types. In Section~\ref{sec:3}, we discuss possible UV completions, starting with all relevant topologies, and then considering the different DM types separately. in Section~\ref{sec:model1}, we construct a simple UV-complete model realization in terms of the well-known scotogenic model and discuss its phenomenology in details. In Section~\ref{sec:5}, we briefly discuss a few other UV-complete model realizations. Our conclusions are given in Section~\ref{sec:6}.

\section{DM-LEFT and DM-SMEFT operators}
\label{sec:2}
To study neutrino-DM interactions in a systematic manner, we first identify the corresponding EFT operators while  keeping the DM field as the only non-SM light degree of freedom in the theory. Within the Low-energy Effective Field Theory (LEFT)~\cite{Buchmuller:1985jz,Jenkins:2017jig} including the DM, which we refer to as DM-LEFT, the relevant DM-LEFT operators are obtained after integrating out massive SM fields such as the $W$, $Z$, Higgs, and top quark, while retaining only the neutrinos and the DM degrees of freedom. We construct all leading-order effective operators that allow neutrino-DM interactions in this framework. We then write the corresponding Standard Model Effective Field Theory (SMEFT) operators~\cite{Grzadkowski:2010es} including the DM field, which we refer to as DM-SMEFT.  At the DM-SMEFT level, we keep the full electroweak gauge symmetry intact, so the DM field couples to the SM lepton ($L$) and Higgs ($H$) doublets. In subsequent sections,  we determine the gauge-invariant UV completions of these DM-SMEFT operators,    and study a few benchmark models that could lead to sizable neutrino-DM interactions while being consistent with theoretical and current experimental constraints. The theoretical constraint arises primarily from neutrino mass induced by the neutrino-DM interactions, while the experimental constraints mainly come from meson decays, as well as from cosmology when the DM mass is below ${\cal O}$(100 MeV).

\begin{table}[t!]
\small
\renewcommand{\arraystretch}{1.3}
    \centering
    \begin{tabular}{|c|l|l|} \hline
        {\bf DM type}  & ~~~~{\bf DM-LEFT operators} & ~~~~~~~~{\bf DM-SMEFT operators}  \\ \hline\hline
        & ${\cal O}_{6,M}^1 \equiv$ $(\nu_L \nu_L) (\chi_L \chi_L)$ & ${\widetilde{\cal O}}_{8,M}^1 \equiv$ $(\vec L\cdot \vec H\ \vec L\cdot \vec H)  (\chi_L \chi_L)$ \\ 
        \cline{2-3}
         Majorana & ${\cal O}_{6,M}^2 \equiv$ $(\nu_L \nu_L) (\chi^c_{~R} \chi^c_{~R})$ & ${\widetilde{\cal O}}_{8,M}^2 \equiv$   $ (\vec L\cdot \vec H\ \vec L\cdot \vec H) \chi^c_{~R} \chi^c_{~R}$ \\ 
        \cline{2-3} 
           Fermion  & \multirow{2}{*}{${\cal O}_{6,M}^3 \equiv$ $(\nu_L \chi_L) (\nu_{~R}^c  \chi^c_{~R}) $} &  ${\widetilde{\cal O}}_{6,M}^3 \equiv (\bar{L} \gamma_\mu L) (\bar{\chi_{L}} \gamma^\mu \chi_{L})$  \\
           & & ${\widetilde{\cal O}}_{8,M}^3 \equiv$ $(\vec L\cdot \vec H \chi_L) (\vec L^c \cdot  \vec{\widetilde{ H}} \chi^c_{~R})$ \\ 
        \cline{2-3}
         $\chi \equiv \chi_L + \chi^c_{~R}$ & ${\cal O}_{6,M}^4 \equiv$ $(\nu_L \chi_L ) (\chi^c_{~R} \chi^c_{~R}) $ & ${\widetilde{\cal O}}_{7,M}^1 \equiv$ $(\vec L\cdot \vec H \chi_L ) (\chi^c_{~R} \chi^c_{~R})  $ \\
        \cline{2-3}
          &  ${\cal O}_{6,M}^5 \equiv$ $(\nu_L \chi_L) (\nu_L \nu_L)$ & ${\widetilde{\cal O}}_{9,M}^1 \equiv$ $(\vec L\cdot \vec H \chi_L ) (\vec L\cdot \vec H\ \vec L\cdot \vec H)  $ \\
        \cline{2-3}
          &   ${\cal O}_{6,M}^6 \equiv$ $(\nu_L \nu_L) (\nu_{~R}^c \chi^c_{~R})$ & ${\widetilde{\cal O}}_{9,M}^2 \equiv$ $(\vec L\cdot \vec H\ \vec L\cdot \vec H) (\vec L^c \cdot  \vec{\widetilde{H}} \chi^c_{~R})  $ \\ \hline \hline
          & \multirow{2}{*}{${\cal O}_{6,D}^1 \equiv$  $(\bar\nu_L \gamma_\mu \nu_L) (\bar\chi_R \gamma^\mu \chi_R) $} &  ${\widetilde{\cal O}}_{6,D}^1 \equiv$  $(\bar L \gamma_\mu L) (\bar \chi_R \gamma^\mu \chi_R)$, \\ 
        & & ${\widetilde{\cal O}}_{8,D}^1 \equiv$  $(H^\dagger \bar L \gamma_\mu \vec L\cdot \vec H) (\bar\chi_R \gamma^\mu \chi_R)$ \\
        \cline{2-3}
         Dirac & \multirow{2}{*}{${\cal O}_{6,D}^2 \equiv$   $(\bar\nu_L \gamma_\mu \nu_L) (\bar\chi_L  \gamma^\mu \chi_L)$} & ${\widetilde{\cal O}}_{6,D}^2 \equiv$  $(\bar L \gamma_\mu L) (\bar \chi_L \gamma_\mu \chi_L)$ \\
         Fermion & & ${\widetilde{\cal O}}_{8,D}^2 \equiv$ $(H^\dagger \bar L \gamma_\mu \vec L\cdot \vec H) (\bar\chi_L \gamma^\mu \chi_L)$ \\
         \cline{2-3}
          $\chi \equiv \chi_L + \chi_R$ & ${\cal O}_{6,D}^3 \equiv$  $(\nu_L \nu_L) (\bar\chi_R \chi_L)$ & ${\widetilde{\cal O}}_{8,D}^3 \equiv$  $(\vec L\cdot \vec H\ \vec L\cdot \vec H) (\bar\chi_R \chi_L)$ \\
          \cline{2-3}
          & ${\cal O}_{6,D}^4 \equiv$  $(\nu_L \nu_L) (\bar\chi_L \chi_R)$ &  ${\widetilde{\cal O}}_{8,D}^4 \equiv$  $(\vec L\cdot \vec H\ \vec L\cdot \vec H) (\bar\chi_L \chi_R)$ \\
          \cline{2-3}
          & ${\cal O}_{6,D}^5 \equiv$  $\left(\nu_{L}\sigma_{\mu \nu} \nu_{L}\right)\left(\bar{\chi}_{R} \sigma^{\mu \nu} \chi_{L}\right) $ & ${\widetilde{\cal O}}_{8,D}^5 \equiv$  $(\vec L\cdot \vec H\ \sigma_{\mu\nu} \vec L\cdot \vec H) (\bar\chi_R \sigma^{\mu\nu} \chi_L)$ \\ \hline\hline
          & ${\cal O}_{4,R}^1 \equiv (\nu_L \nu_L)\ \Phi$ & ${\cal \widetilde{O}}_{6,R}^1 \equiv (\vec L \cdot \vec H\ \vec L \cdot \vec H)\ \Phi$ \\ 
         \cline{2-3}
          & ${\cal O}_{5,R}^1 \equiv (\nu_L \nu_L)\ \Phi \Phi$ &  ${\cal \widetilde{O}}_{7,R}^1 \equiv (\vec L \cdot \vec H\ \vec L \cdot \vec H)\ \Phi \Phi$ \\
          \cline{2-3}
         Real Scalar & \multirow{4}{*}{${\cal O}_{6,R}^1 \equiv (\nu_L \sigma_{\mu\nu}\nu_L)\ \Phi F^{\mu\nu}$} & ${\cal \widetilde{O}}_{8,R}^1 \equiv (\vec L \cdot \vec H\ \sigma_{\mu\nu}\vec L \cdot \vec H)\ \Phi B^{\mu\nu}$,\ \\
          $\Phi$ & & ${\cal \widetilde{O}}_{8,R}^{1^\prime} \equiv (\vec L \cdot \vec H\ \sigma_{\mu\nu}\tau^I \vec L \cdot \vec H)\ \Phi W^{I,\mu\nu}$\\
          & & ${\cal \widetilde{O}}_{7,R}^2 \equiv \bar L \gamma_\mu L\ (D_\nu \Phi) B^{\mu \nu}$, \\
          & & ${\cal \widetilde{O}}_{7,R}^{2^\prime} \equiv \bar L \gamma_\mu \tau^I L\ (D_\nu \Phi) W^{I \mu \nu}$ \\
          \cline{2-3}
          & ${\cal O}_{6,R}^2 \equiv (\nu_L \nu_L)\ \Phi \Phi \Phi$ & ${\cal \widetilde{O}}_{8,R}^2 \equiv (\vec L \cdot \vec H\ \vec L \cdot \vec H)\ \Phi \Phi \Phi$ \\ \hline\hline
         Pseudoscalar & \multirow{2}{*}{${\cal O}_{5,P}^1 \equiv (\bar\nu_L \gamma_\mu \nu_L) \partial^\mu \Phi$} & ${\cal \widetilde{O}}_{5,P}^1 \equiv (\bar L \gamma_\mu L)\ D^\mu \Phi$\\
         $\Phi$ & & ${\cal \widetilde{O}}_{7,P}^1 \equiv ( H^\dagger \bar L\ \gamma_\mu \vec L \cdot \vec H)\ D^\mu \Phi$ \\ \hline\hline
       Complex scalar  & ${\cal O}_{5,C}^1 \equiv (\nu_L \nu_L)\ \Phi^\dagger \Phi$ &  ${\cal \widetilde{O}}_{7,C}^1 \equiv (\vec L \cdot \vec H\ \vec L \cdot \vec H)\ \Phi^\dagger \Phi$ \\ 
        \cline{2-3}
       $\Phi$ & \multirow{2}{*}{${\cal O}_{6,C}^1 \equiv (\bar\nu_L \gamma_\mu \nu_L)\ (\partial^\mu \Phi^\dagger) \Phi$} &  ${\cal \widetilde{O}}_{6,C}^1 \equiv (\bar L \gamma_\mu L)\ (\partial^\mu \Phi^\dagger) \Phi$  \\  
       & & ${\cal \widetilde{O}}_{8,C}^1 \equiv (H^\dagger \bar L \gamma_\mu \vec L \cdot \vec H)\ (\partial^\mu \Phi^\dagger) \Phi$ \\ \hline\hline
       &  \multirow{2}{*}{${\cal O}_{4V}^1 \equiv (\bar \nu_L \gamma_\mu \nu_L) \chi^\mu$} &  ${\widetilde{\cal O}}_{4V}^1 \equiv (\bar L \gamma_\mu L) \chi^\mu$  \\
      & & ${\widetilde{\cal O}}_{6V}^1 \equiv (H^\dagger \bar L \gamma_\mu \vec L \cdot \vec H) \chi^\mu$ \\
      \cline{2-3}
      Vector & ${\cal O}_{5V}^1 \equiv (\nu_L \sigma_{\mu\nu} \nu_L) \chi^{\mu\nu}$ & ${\widetilde{\cal O}}_{7V}^1 \equiv (\vec L \cdot \vec H \sigma_{\mu\nu}  \vec L \cdot \vec H) \chi^{\mu\nu}$ \\
       \cline{2-3}
     Boson  & ${\cal O}_{5V}^2 \equiv (\nu_L \nu_L) \chi_\mu \chi^{\mu}$ & ${\widetilde{\cal O}}_{7V}^2 \equiv (\vec L \cdot \vec H\  \vec L \cdot \vec H) \chi_{\mu} \chi^\mu$ \\
       \cline{2-3}
       $\chi^\mu$ & ${\cal O}_{5V}^3 \equiv (\nu_L \nu_L) \partial_\mu \chi^{\mu}$ &  ${\widetilde{\cal O}}_{7V}^3 \equiv (\vec L \cdot \vec H\  \vec L \cdot \vec H) \partial_{\mu} \chi^\mu$ \\
       \cline{2-3}
       & ${\cal O}_{5V}^4 \equiv (\nu_L \overleftrightarrow{\partial_\mu} \nu_L) \chi^{\mu}$ & ${\widetilde{\cal O}}_{7V}^4 \equiv (\vec L \cdot \vec H\ \overleftrightarrow{D_\mu}  \vec L \cdot \vec H) \chi^\mu$\\ \hline
    \end{tabular}
    \caption{The relevant {\bf DM-LEFT and DM-SMEFT operators} for different DM types, namely, Majorana/Dirac fermion, real/pseudo/complex scalar, and vector DM. Here ``$\cdot$" represents $SU(2)$ contraction with antisymmetric tensor $\epsilon_{ij}$; $\chi^{\mu\nu}$, $F^{\mu\nu}$, $B^{\mu\nu}$, $W^{\mu\nu}$ are the DM, electromagnetic, $U(1)_Y$, and $SU(2)_L$ field strength tensors, respectively, and $D^\mu$ is the covariant derivative. The notation $\chi^c_{~R}$ (and similarly $\nu^c_{~R}$) is adopted to explicitly indicate that it  is a right-chiral field. We use the four-component spinor notation throughout.
    }
    \label{tab:operators}
\end{table}

For fermionic DM, we include all non–derivative interactions\footnote{Derivative operators are not part of the minimal basis, since they can generally be eliminated using equations of motion (EOM) and/or  integration by parts. For example, for a real scalar DM field~$\Phi$, the operator $(\bar\nu_L \gamma^\mu \nu_L)(\partial_\mu \Phi)\Phi$ can be reduced to a redundant basis using the EOM of~$\Phi$. } up to dimension-6 at the DM-LEFT level, which correspond to operators up to dimension-8 in the DM-SMEFT. For scalar and
vector DM, the relevant non-derivative interactions already appear at dimension-5 in DM-LEFT and up to dimension-7 in DM-SMEFT. Table~\ref{tab:operators} summarizes the independent operator structures that mediate interactions between neutrinos and different types of DM candidates.  For each DM spin and Lorentz structure, we list the leading DM-LEFT operators and their corresponding gauge–invariant DM-SMEFT counterparts.   
Electroweak gauge invariance fixes $SU(2)_L$ contractions uniquely and we use the notation 
\begin{equation}
    \vec{L}\!\cdot\!\vec{H} \equiv \epsilon_{ij} L^i H^j \, ,
\end{equation}
with $\epsilon_{ij}$ being the $SU(2)$ anti-symmetric tensor. This allows the operators to be expressed compactly in terms of singlet contractions $(\vec{L}\!\cdot\!\vec{H})$ and $(\vec{L}^{\,c}\!\cdot\!\vec{\widetilde{H}})$,  which after electroweak symmetry breaking (EWSB) yield operators involving the neutrino fields~$\nu_L$ appearing in the DM-LEFT. Here $\overline{L^c} \equiv L^T C$, with $C$ being the charge conjugation operator, and  $\widetilde H \equiv i\tau_2 H^*$, with $\tau_2$ being the second Pauli matrix. We use the four-component spinor notation throughout the text. For instance, $(\nu_L\nu_L)\equiv \overline{(\nu_L)^c}\nu_L=\overline{{\nu^c}_R}\nu_L=\nu_L^TC\nu_L$. 

Schematically, the two DM-LEFT and DM-SMEFT operators are related by
\begin{equation}
  \mathcal{O}^{\rm DM-LEFT}_{d}
  = \frac{v^{n}}{\Lambda^{n}}\,
  \widetilde{\mathcal{O}}^{\rm DM-SMEFT}_{d+n}\,,
\end{equation}
where $v\simeq 246.2~{\rm GeV}$ is the electroweak vacuum expectation value (VEV), $d$ denotes the operator dimension in DM-LEFT, $n$ counts the number of Higgs fields that acquire VEVs after EWSB, and $\Lambda$ is the new physics scale above the electroweak scale.


Some of the operators in Table~\ref{tab:operators}, particularly those containing $(\vec L \cdot \vec H\ \vec L \cdot \vec H )$ structure, may violate lepton number by two units ($\Delta L = 2$) and therefore can generate Majorana neutrino masses once the DM field is integrated out~\cite{Weinberg:1979sa,Babu:2001ex}. In many EFT operators listed in Table~\ref{tab:operators}, the DM bilinear (e.g., $\chi_L \chi_L$ or $\Phi \Phi$) can be closed in a loop, inducing radiative neutrino masses~\cite{Bonnet:2012kz,Cai:2017jrq} through the same effective coupling that mediates neutrino–DM scattering. A naive one-loop estimate for the induced Majorana neutrino mass is given by
\begin{equation}
  m_\nu \;\sim\; \frac{C} {16\pi^2}\; \frac{m_{\rm DM} v^2}{\Lambda^2} \,,
  \label{eq:mnu_LEFT}
\end{equation}
where $C$ is some dimensionless LEFT Wilson coefficient and $\Lambda$ is the scale of the heavy mediator. This naive link between neutrino-DM couplings and neutrino mass is, of course, model-dependent, as it can be weakened or modified by flavor structure, loop order, masses of the heavy states, and parameters that generate both the neutrino-mass operator and the DM–neutrino interaction. For details, see the specific model examples in the next sections. 
Some additional remarks related to Table~\ref{tab:operators} are in order:
\begin{itemize}
  \item There are other invariants analogous to ${\widetilde{\cal O}}_{8,M}^3$ , ${\widetilde{\cal O}}_{8,D}^1$, and  ${\widetilde{\cal O}}_{8,D}^2$. However, they do not generate the $2\nu2\chi$ DM-LEFT operators and thus UV completion of such operators is beyond the scope of current work. An example is $(\vec L\cdot \vec{\widetilde{ H}} \chi_L) (\vec L^c \cdot \vec H   \chi^c_{~R})$.
  \item There are other dimension-6 operators for vector DM-SMEFT: ${\widetilde{\cal O}}_{6V}^1 \equiv (\bar L \gamma_\mu L) \chi^\mu \partial_\nu \chi^\nu$, ${\widetilde{\cal O}}_{6V}^2 \equiv (\bar L \gamma_\mu L) \chi_\nu  \chi^{\mu\nu}$,  ${\widetilde{\cal O}}_{6V}^3 \equiv (\bar L \gamma_\mu L) \chi_\nu  B^{\mu\nu}$, and ${\widetilde{\cal O}}_{6V}^4 \equiv (\bar L \gamma_\mu \tau^I L) \chi_\nu  W^{I\mu\nu}$. The corresponding DM-LEFT operators which are at the same dimension are obtained by replacing $L$ by $\nu_L$. Since all of them contain derivatives, they are omitted from the list in Table~\ref{tab:operators}. 
  \item For a Majorana fermion DM, the vector bilinear vanishes ($\bar\chi\gamma_\mu\chi=0$), while
  the axial vector combination ($\bar\chi\gamma_\mu\gamma_5\chi\neq 0$) survives. Thus, only the axial current contributes to the operator $\widetilde{\mathcal{O}}_{6,M}^{3}$.
\end{itemize}
In summary, the operators shown in Table~\ref{tab:operators} constitute a  non–redundant set of the lowest–dimensional interactions that couple neutrinos to DM within the DM-LEFT and DM-SMEFT frameworks. These operators provide the starting point for identifying renormalizable UV completions, which we analyze systematically in the next section.

\section{UV completions}
\label{sec:3}

EFT descriptions are intentionally agnostic about the UV dynamics, since  different UV completions can match onto the same EFT operator basis, and conversely a given set of heavy mediators can generate several linearly independent EFT operators. In this work, we focus on the simplest realization, namely renormalizable tree-level completions of the DM-SMEFT operators. It is then straightforward to draw the Feynman diagrams or topologies at renormalizable level that will lead to such effective operators. For a given operator, we only require topologies with the same number of external fermion and boson legs as the corresponding DM-SMEFT interaction. Moreover, we restrict to inequivalent (non-isomorphic) topologies, since all distinct attachments of the operator fields to the external legs are generated by permutations. 

To automate this classification, we use a publicly available code~\cite{Heeck:2026dmh} interfaced with the Mathematica package {\tt IGraph/M}~\cite{Horv_t_2023}, together with {\tt FeynArts}~\cite{Kublbeck:1990xc}, to generate all non-isomorphic topologies with the appropriate external-leg structure. For each non-derivative DM-SMEFT operator of mass dimension $d$, the algorithm assigns the operator fields to the external legs and determines the allowed gauge quantum numbers of the internal mediators by successive multiplication of $SU(3)_c\times SU(2)_L\times U(1)_Y$ representations using \texttt{GroupMath}~\cite{Fonseca:2020vke}. We take internal fermions to be Dirac fields, and we also require a mass insertion along each internal fermion line, which ensures that derivative operators are not generated~\cite{Gargalionis:2020xvt}. For completeness, we present an explicit example of derivative UV completion in the case of scalar DM in section~\ref{sec:scalardm}. 

We note that the UV completions may, in principle, contain fields that are lighter than the electroweak scale. In such scenarios, the corresponding DM-SMEFT operators would be modified by the presence of light degrees of freedom. For simplicity, we do not explicitly write these modified operators. Instead, when matching onto the low-energy theory, we assume that those fields are heavier than the DM mass and can therefore be integrated out, yielding the DM-LEFT operators of Table~\ref{tab:operators}. 

\subsection{Relevant topologies}
\begin{figure}[htb!]
    \centering
    \begin{subfigure}{0.9\linewidth}
        \centering
        \includegraphics[width=\linewidth]{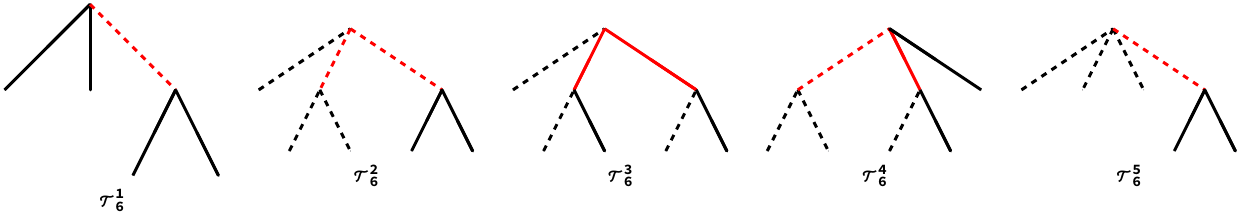}
        \caption{Tree-level topologies relevant for $d=6$ DM-SMEFT operators.}
        \label{fig:dim6}
    \end{subfigure}
\vspace{3mm}
    \begin{subfigure}{0.9\linewidth}
        \centering
        \includegraphics[width=\linewidth]{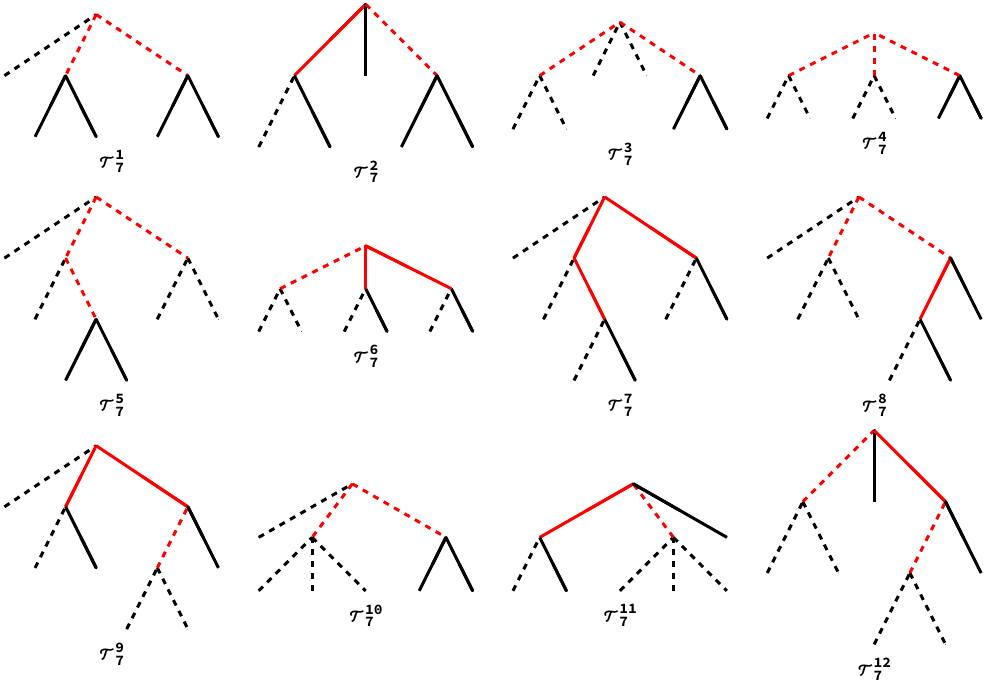}
        \caption{Tree-level renormalizable topologies relevant for $d=7$ DM-SMEFT operators.}
        \label{fig:dim7}
    \end{subfigure}
\vspace{3mm}
    \begin{subfigure}{0.9\linewidth}
        \centering    \includegraphics[width=\linewidth]{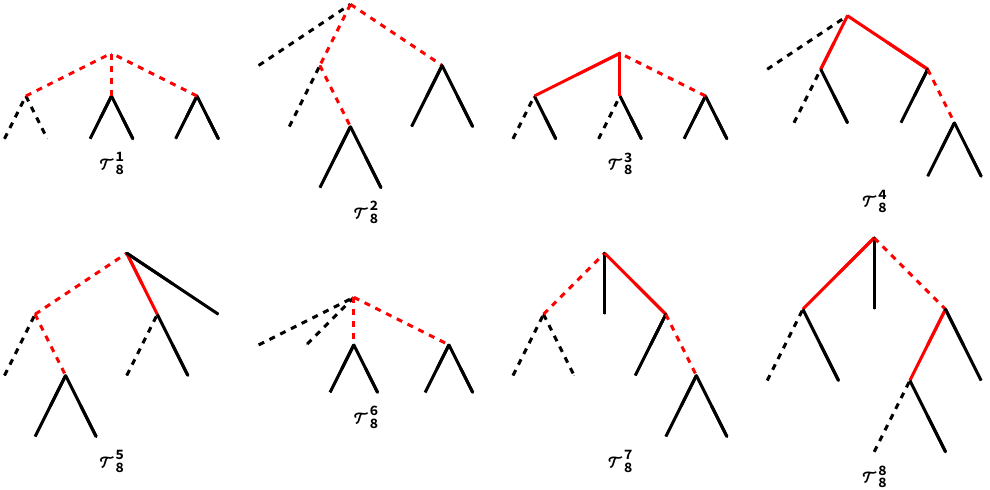}
        \caption{Tree-level renormalizable topologies relevant for $d=8$ DM-SMEFT operators.}
        \label{fig:dim8}
    \end{subfigure}
    \caption{Tree-level renormalizable diagram topologies used to construct UV completions of the DM-SMEFT operators in Table~\ref{tab:operators} at $d=6$, 7 and 8. External legs (black) represent the operator fields ($L$, $H$, and the dark-sector fields), while internal lines (red) denote heavy mediators. Solid (dashed) lines correspond to fermions (bosons).}
    \label{fig:dim678}
\end{figure}

We now construct the tree-level renormalizable diagram topologies (as shown in Figs.~\ref{fig:dim6}, \ref{fig:dim7}, and \ref{fig:dim8}) that can UV-complete the non-derivative DM-SMEFT operators listed in Table~\ref{tab:operators}. Our focus is on operators of mass dimension $d=6$-$8$, for which we generate all connected, non-isomorphic topologies with the appropriate external field content. It is important to note that for vector DM,  there is also a single $d=4$ operator in Table~\ref{tab:operators}, which we treat separately. At $d=5$ one may write derivative operators, but within the operator set considered here there is no $d=5$ DM-SMEFT interaction that matches onto the required non-derivative DM-LEFT operators governing neutrino-DM scattering. 

In each diagram the external legs correspond to the fields appearing in the DM-SMEFT operators, namely the SM lepton doublet $L$, the Higgs doublet $H$, and the dark-sector fields (fermionic $\chi$, scalar $\Phi$, and vector $\chi^\mu$)--and are drawn in black. Internal lines denote heavy mediators and are shown in red. Solid lines represent fermions, while dashed lines denote bosons (scalars or vectors).  Since permutations of the operator fields over  external legs
\clearpage
\noindent
are treated explicitly in the operator assignment step, each topology shown below represents an equivalence class of diagrams up to relabeling of the external legs.

The resulting irreducible topologies for $d=6$-$8$ are displayed in Figs.~\ref{fig:dim6}-\ref{fig:dim8}. For $d=6$ (Fig.~\ref{fig:dim6}) we obtain two distinct external-leg structures relevant to Table~\ref{tab:operators}: (i) four-fermion topologies, denoted ${\cal T}_6^1$, and (ii) topologies with two external fermions and three external scalars, denoted ${\cal T}_6^2$-${\cal T}_6^5$. 
For $d=7$ (Fig.~\ref{fig:dim7}) the corresponding classes are: (i) four fermions plus one scalar, ${\cal T}_7^{1,2}$, and (ii) two fermions plus four scalars, ${\cal T}_7^{3}$-${\cal T}_7^{12}$. 
Finally, for the $d=8$ operator set considered here (Fig.~\ref{fig:dim8}), all irreducible tree-level topologies involve four external fermions and two external scalars.

\begin{table}[!t]
    \centering
    \begin{tabular}{|c|c|}\hline
       {\bf Scalars}  & $\phi_1 (1,1,0)$,\hspace{1mm} $\phi_2 (1,1,1)$,\hspace{1mm} $\phi_3 (1,2,1/2)$,\hspace{1mm} $\phi_4 (1,3,0)$,\hspace{1mm} $\phi_5 (1,3,1)$  \\ \hline
       {\bf Fermions}  &  $\psi_1 (1,1,0)$,\hspace{1mm} $\psi_2 (1,1,1)$,\hspace{1mm} $\psi_3 (1,2,1/2)$,\hspace{1mm} $\psi_4 (1,3,0)$,\hspace{1mm} $\psi_5 (1,3,1)$ \\ \hline 
    \end{tabular}
    \caption{Particle representations of new Dirac fermions $\psi_i \equiv \psi_{iL} + \psi_{iR}$ and scalars $\phi_i$ under the SM gauge group $SU(3)_c \times SU(2)_L \times U(1)_Y$ that mediate the neutrino-DM interactions and generate the operators of Table~\ref{tab:operators} after being integrated out.}
    \label{tab:names}
\end{table}

We assign names of the new scalar and vector-like (Dirac) fermion fields in Table~\ref{tab:names}, classified under the SM gauge group $SU(3)_c \times SU(2)_L \times U(1)_Y$. Table~\ref{tab:names} contains mediators that transforms as singlets, doublets, or triplets of $SU(2)_L$. This restriction arises from the requirement of renormalizable interactions involving the SM lepton ($L$) and Higgs ($H$) doublets and SM-singlet DM fields. Consequently, mediators in larger $SU(2)_L$-representations are not generated in tree-level UV completions of the DM-SMEFT operators considered here. 

Throughout our analysis, we treat internal mediators as distinguishable, even when they share identical gauge quantum numbers.  In many cases, a single copy of a mediator field suffices, whereas certain realizations involving antisymmetric contractions require multiple copies to obtain a non-vanishing amplitude. When constructing the minimal benchmark models for sizable neutrino-DM interactions in the subsequent sections, we retain only the smallest mediator content needed for the desired operator(s). 

In the following subsections, we construct  explicit UV completions for different DM types. 
\subsection{Majorana fermion DM}
\newcolumntype{C}{>{$}l<{$}}
\begin{table}[!h]
\small
\renewcommand{\arraystretch}{1.1}
    \centering
    \begin{tabular}{|C|C|C|}
    \hline
     
   {\rm \bf Operator}   & {\rm  \bf Topology}  & {\rm  \bf \hspace{4cm} New\ Particles}  \\
        \hline\hline
   \multirow{8}{*}{\rotatebox{90}{\({\widetilde{\cal O}}_{8,M}^1 \equiv L^2H^2\chi_L\chi_L\)}}   & {\cal T}_8^1  & \{\phi _5 \phi _5 \phi _1\},\ \{\phi _5 \phi _3 \phi _3\} \\
         \cline{2-3}
      & {\cal T}_8^2  & \{\phi _1 \phi _3 \phi _5\},\ \{\phi _1 \phi _3 \phi _2\},\ \{\phi _3 \phi _4 \phi _3\},\ \{\phi _3 \phi _1 \phi _3\} \\
       \cline{2-3}
      & {\cal T}_8^3  &\{\psi _4 \psi _4 \phi _1\},\ \{\psi _1 \psi _1 \phi _1\},\ \{\psi _4 \psi _3 \phi _3\},\  \{\psi _1 \psi _3 \phi _3\} \\
       \cline{2-3}
      & {\cal T}_8^4  &  \{\phi _1 \psi _3 \psi _4\},\  \{\phi _1 \psi _3 \psi _1\},\ \{\phi _3 \psi _5 \psi _3\},\ \{\phi _3 \psi _2 \psi _3\}\\
       \cline{2-3}
      & {\cal T}_8^5  & \{\psi _3 \phi _4 \phi _3\},\ \{\psi _3 \phi _1 \phi _3\},\ \{\psi _4 \phi _3 \phi _1\},\ \{\psi _1 \phi _3 \phi _1\} \\
       \cline{2-3}
      & {\cal T}_8^6  & \{\phi _5 \phi _1\},\  \{\phi _3 \phi _3\} \\
       \cline{2-3}
      & {\cal T}_8^7  & \{\phi _1 \psi _3 \phi _5\},\ \{\phi _3 \psi _3 \phi _5\},\ \{\phi _3 \psi _5 \phi _5\},\  \{\phi _5 \psi _5 \phi _5\} \\
       \cline{2-3}
      & {\cal T}_8^8  & \{\psi _3 \phi _4 \psi _3\},\ \{\psi _3 \phi _1 \psi _3\},\ \{\psi _3 \phi _3 \psi _4\},\ \{\psi _3 \phi _3 \psi _1\},\ \{\psi _4 \phi _4 \psi _3\},\ \{\psi _1 \phi _1 \psi _3\}  \\
       \hline \hline
        \multirow{8}{*}{\rotatebox{90}{\({\widetilde{\cal O}}_{8,M}^2 \equiv  L^2 H^2 \chi^c_{~R} \chi^c_{~R}\)}} 
       & {\cal T}_8^1 & \{\phi _5 \phi _5 \phi _1\} \\ \cline{2-3}
       & {\cal T}_8^2& \{\phi _1 \phi _3 \phi _5\},\ \{\phi _1 \phi _3 \phi _2\} \\ \cline{2-3}
       &{\cal T}_8^3& \{\psi _4 \psi _4 \phi _1\},\ \{\psi _1 \psi _1 \phi _1\} \\ \cline{2-3}
       &{\cal T}_8^4 & \{\phi _1 \psi _3 \psi _4\},\ \{\phi _1 \psi _3 \psi _1\}  \\ \cline{2-3}
       &{\cal T}_8^5 &  \{\psi _4 \phi _3 \phi _1\},\ \{\psi _1 \phi _3 \phi _1\}\\ \cline{2-3}
       &{\cal T}_8^6 & \{\phi _5 \phi _1\} \\ \cline{2-3}
       & {\cal T}_8^7 & \{\phi _1 \psi _3 \phi _5\},\ \{\phi _5 \psi _5 \phi _5\} \\ \cline{2-3}
       & {\cal T}_8^8 & \{\psi _3 \phi _3 \psi _4\},\ \{\psi _3 \phi _3 \psi _1\} \\ \hline \hline
       \multirow{15}{*}{\rotatebox{90}{\({\widetilde{\cal O}}_{8,M}^3 \equiv LL^c H \widetilde{H}\chi_L \chi^c_{~R}\)}} 
     & {\cal T}_8^1 & \{\phi _4 \phi _3 \phi _3\},\ \{\phi _1 \phi _3 \phi _3\} \\  \cline{2-3} 
     & {\cal T}_8^2 & \{\phi _3 \phi _4 \phi _3\},\ \{\phi _3 \phi _1 \phi _3\},\ \{\phi _3 \phi _5 \phi _3\},\ \{\phi _3 \phi _2 \phi _3\}\\ \cline{2-3}
     & \multirow{2}{*}{${\cal T}_8^3$} & \{\psi _4 \psi _3 \phi _3\},\ \{\psi _1 \psi _3 \phi _3\},\ \{\psi _5 \psi _3 \phi _3\},\ \{\psi _2 \psi _3 \phi _3\},\ \{\psi _3 \psi _5 \phi _3\},\ \{\psi _3 \psi _2 \phi _3\}, \\ 
     & &  \{\psi _3 \psi _4 \phi _3\},\ \{\psi _3 \psi _1 \phi _3\} \\ \cline{2-3}
     & \multirow{2}{*}{${\cal T}_8^4$} & \{\phi _3 \psi _4 \psi _3\},\ \{\phi _3 \psi _1 \psi _3\},\ \{\phi _3 \psi _4 \psi _3\},\ \{\phi _3 \psi _1 \psi _3\},\ \{\phi _3 \psi _3 \psi _5\},\ \{\phi _3 \psi _3 \psi _2\}, \\ 
     & & \{\phi _3 \psi _3 \psi _5\},\ \{\phi _3 \psi _3 \psi _2\} \\ \cline{2-3}
     & \multirow{3}{*}{${\cal T}_8^5$} & \{\psi _3 \phi _5 \phi _3\},\ \{\psi _3 \phi _2 \phi _3\},\ \{\psi _3 \phi _4 \phi _3\},\ \{\psi _3 \phi _1 \phi _3\},\ \{\psi _5 \phi _5 \phi _3\},\  \{\psi _2 \phi _2 \phi _3\}, \\ 
     & &  \{\psi _4 \phi _4 \phi _3\},\ \{\psi _1 \phi _1 \phi _3\},\ \{\psi _3 \phi _4 \phi _3\},\ \{\psi _3 \phi _1 \phi _3\},\ \{\psi _3 \phi _5 \phi _3\},\ \{\psi _3 \phi _2 \phi _3\},\\ 
     & & \{\psi _4 \phi _4 \phi _3\},\ \{\psi _1 \phi _1 \phi _3\},\ \{\psi _5 \phi _5 \phi _3\},\ \{\psi _2 \phi _2 \phi _3\}\\ \cline{2-3}
     & {\cal T}_8^6 & \{\phi _3 \phi _3\} \\ \cline{2-3}
     & \multirow{2}{*}{${\cal T}_8^7$} & \{\phi _3 \psi _3 \phi _4\},\ \{\phi _3 \psi _3 \phi _1\},\ \{\phi _3 \psi _3 \phi _4\},\ \{\phi _3 \psi _3 \phi _1\},\ \{\phi _3 \psi _4 \phi _4\},\ \{\phi _3 \psi _1 \phi _1\}, \\ 
     & &  \{\phi _3 \psi _4 \phi _4\},\ \{\phi _3 \psi _1 \phi _1\}\\ \cline{2-3} 
    &  \multirow{3}{*}{${\cal T}_8^8$} & \{\psi _3 \phi _4 \psi _3\},\ \{\psi _3 \phi _1 \psi _3\},\  \{\psi _4 \phi _4 \psi _3\},\ \{\psi _1 \phi _1 \psi _3\},\  \{\psi _3 \phi _5 \psi _3\},\ \{\psi _3 \phi _2 \psi _3\}, \\ 
     & & \{\psi _5 \phi _5 \psi _3\},\ \{\psi _2 \phi _2 \psi _3\},\ \{\psi _3 \phi _4 \psi _4\},\ \{\psi _3 \phi _1 \psi _1\},\ \{\psi _4 \phi _4 \psi _4\},\ \{\psi _1 \phi _1 \psi _1\},\\ 
     & & \{\psi _3 \phi _5 \psi _5\},\ \{\psi _3 \phi _2 \psi _2\},\ \{\psi _5 \phi _5 \psi _5\},\ \{\psi _2 \phi _2 \psi _2\}\\ \hline \hline 
      \multirow{2}{*}{\rotatebox{90}{\({\widetilde{\cal O}}_{7,M}^1\)}} 
       & {\cal T}_7^1 & \{\phi _3 \phi _1\} \\ \cline{2-3}
       & {\cal T}_7^2& \{\psi _3 \phi _1\},\ \{\psi _1 \phi _1\},\ \{\psi _3 \phi _3\} \\ \hline \hline 
      {\widetilde{\cal O}}_{6,M}^3
       & {\cal T}_6^1 & \{\phi _3 \} \\ \hline
    \end{tabular}
    \caption{Majorana DM UV completions. The first column lists the relevant DM-SMEFT operator, the second column the irreducible topologies, while the last column specifies representative sets of new heavy fermions ($\psi_i$) and scalars ($\phi_i$). }
    \label{tab:MajUV}
\end{table}
We start with the case of Majorana fermionic DM, defined as $\chi \equiv \chi_L + \chi^c_{~R} $ 
and taken to be SM-singlet with quantum numbers $(1,1,0)$ under $SU(3)_c \times SU(2)_L \times U(1)_Y$. In this section, we remain agnostic about the additional symmetries (e.g.\ a $\mathbb{Z}_2$-parity or a remnant of
a broken gauge symmetry) that render $\chi$ a viable DM candidate. Such symmetries 
will be made explicit for the concrete UV models presented in subsequent sections. The Majorana mass term is given by $\tfrac{1}{2} m_\chi\, \chi^T C \chi$. 
Throughout this work we focus on
$m_\chi \lesssim v$, i.e.\ DM mass below the electroweak scale, so that $\chi$ remains an explicit dynamical degree of freedom both above and below the EWSB. In this regime the DM-SMEFT operators match onto DM-LEFT interactions that directly control low-energy phenomenology.

At dimension-6, the relevant DM-LEFT operator basis for Majorana $\chi$ is given by ${\cal O}^1_{6,M}$-${\cal O}^6_{6,M}$, listed in the second column of Table~\ref{tab:operators}, together with their gauge-invariant DM-SMEFT embeddings shown in the third column. In this section, we construct explicit renormalizable UV completions for the following (schematically written) operators:
\begin{align}
&\widetilde{\cal O}_{8,M}^1 \equiv L^2 H^2\, \chi_L \chi_L, \hspace{15mm}
\widetilde{\cal O}_{8,M}^2 \equiv L^2 H^2\, \chi^c_{~R} \chi^c_{~R},  \hspace{10mm} 
\widetilde{\cal O}_{8,M}^3 \equiv L\, L^c\, H\, \widetilde H\, \chi_L \chi^c_{~R}, \notag \\ 
&\widetilde{\cal O}_{7,M}^1 \equiv L\,H\, \chi_L \chi^c_{~R} \chi^c_{~R}\, , \hspace{10mm} {\widetilde{\cal O}}_{6,M}^3 \equiv (\bar{L} \gamma_\mu L) (\bar{\chi_{L}} \gamma^\mu \chi_{L}).
\label{eq:Majopt}
\end{align}
Here $L^2$ and $H^2$ indicate two insertions of the corresponding SM doublets, and we suppress Lorentz and gauge-index contractions for brevity. In practice, our topology-based UV-completion procedure automatically accounts for all allowed contractions and all inequivalent attachments of the external DM-SMEFT fields to a given diagram topology.

For simplicity, we do not pursue UV completions of the dimension-9 operators $\widetilde{\cal O}_{9,M}^1$ and $\widetilde{\cal O}_{9,M}^2$. Besides requiring a substantially larger set of topologies (four external fermions plus three external scalars), these operators yield, after EWSB, a low-energy interaction of the schematic form $C_{\rm eff}\,\chi\,\nu\,\nu\,\nu$, which mediates dark-matter decay and can therefore jeopardize the stability of $\chi$. Imposing the current bound on the lifetime of an ${\cal O}(100~{\rm MeV})$ DM decaying to neutrinos,  $\tau_\chi\gtrsim 10^{23}$ s~\cite{Arguelles:2022nbl}, leads to the approximate constraint
\begin{equation}
  |C_{\rm eff}|
  \;\lesssim\;
  6.3\times10^{-20}\ \text{GeV}^{-2}\,
  \left(\frac{100~\text{MeV}}{m_\chi}\right)^{5/2},
  \label{eq:Ceff-bound}
\end{equation}
where $C_{\rm eff}\equiv C\,v^3/\Lambda^5$ denotes the effective LEFT coupling induced by a dimension-9 DM-SMEFT operator with Wilson coefficient $C$ and cutoff scale $\Lambda$. 

The allowed tree-level renormalizable UV completions of the operators in Eq.~\eqref{eq:Majopt}, together with their associated irreducible diagram topologies, are summarized in Table~\ref{tab:MajUV}. The second column lists the topology class, while the last column gives representative choices of heavy fermion ($\psi$) and scalar ($\phi$) mediators that realize the corresponding effective interactions. Curly braces indicate the set of heavy fields propagating internally for a given topology. We also display cases involving repeated mediator representations.  Although such assignments may correspond to fields carrying identical gauge quantum numbers, certain contractions vanish due to antisymmetry, and in those instances multiple copies (or distinguishable states) are required to obtain a non-zero amplitude.

\subsection{Dirac fermion DM}
In this subsection, we discuss the realization of fermionic DM of Dirac
type, defined as $\chi \equiv \chi_L + \chi_R$ 
taken to be SM-singlet with quantum numbers $(1,1,0)$ under the SM gauge group. Unlike the Majorana case, here we assume that $\chi$ is charged under an additional global $U(1)$ symmetry with charge $q_\chi$, transforming as
\begin{equation}
    \chi \;\to\; e^{i q_\chi \alpha}\, \chi \, ,
\end{equation}
where $\alpha$ is the global phase parameter.
The Dirac mass term
\begin{equation}
   - \mathcal{L} \supset m_\chi\, \bar{\chi}_L \chi_R + \text{H.c.}\, 
\end{equation}
is invariant under this $U(1)$, provided the two chiralities carry the same charge, 
$q_{\chi_L} = q_{\chi_R} \equiv q_\chi$. Since the UV completions must respect this global symmetry, the heavy mediator fields 
\begin{table}[!h]
\renewcommand{\arraystretch}{1.27}
\small
    \centering
    \begin{tabular}{|C|C|C|}
    \hline
     {\rm \bf Operator} & {\rm \bf Topology} & {\rm  \bf \hspace{4cm} New\ Particles}  \\
        \hline\hline
   {\widetilde{\cal O}}_{6,D}^1    & {\cal T}_6^1  & \{\widetilde{\phi }_3\} \\ \hline \hline
       \multirow{13}{*}{\rotatebox{90}{\({\widetilde{\cal O}}_{8,D}^1 \equiv \bar L L H^\dagger H \bar\chi_R \chi_R\)}} 
      & {\cal T}_8^1  & \{\phi _4 \widetilde{\phi }_3 \widetilde{\phi }_3\},\ \{\phi _1 \widetilde{\phi }_3 \widetilde{\phi }_3\} \\ \cline{2-3}
      & {\cal T}_8^2 & \{\widetilde{\phi }_3 \widetilde{\phi }_4 \widetilde{\phi }_3\},\ \{\widetilde{\phi }_3 \widetilde{\phi }_1 \widetilde{\phi }_3\},\ \{\widetilde{\phi }_3 \widetilde{\phi }_5 \widetilde{\phi }_3\},\ \{\widetilde{\phi }_3 \widetilde{\phi }_2 \widetilde{\phi }_3\}\ \\ \cline{2-3}
      & \multirow{2}{*}{${\cal T}_8^3$} & \{\psi _4 \widetilde{\psi }_3 \widetilde{\phi }_3\},\ \{\psi _1 \widetilde{\psi }_3 \widetilde{\phi }_3\},\ \{\psi _5 \widetilde{\psi }_3^' \widetilde{\phi }_3\},\ \{\psi _2 \widetilde{\psi }_3^' \widetilde{\phi }_3\},\ \{\widetilde{\psi }_3^' \psi _5 \widetilde{\phi }_3\},\ \{\widetilde{\psi }_3^' \psi _2 \widetilde{\phi }_3\},\ \{\widetilde{\psi }_3 \psi _4 \widetilde{\phi }_3\}, \\
      & & \{\widetilde{\psi }_3 \psi _1 \widetilde{\phi }_3\}\\ \cline{2-3}
      & \multirow{2}{*}{${\cal T}_8^4$} & \{\widetilde{\phi }_3 \widetilde{\psi }_4 \widetilde{\psi }_3\},\ \{\widetilde{\phi }_3 \widetilde{\psi }_1 \widetilde{\psi }_3\},\ \{\widetilde{\phi }_3 \widetilde{\psi }_4 \widetilde{\psi }_3^'\},\ \{\widetilde{\phi }_3 \widetilde{\psi }_1 \widetilde{\psi }_3^'\},\ \{\widetilde{\phi }_3 \psi _3 \psi _5\},\ \{\widetilde{\phi }_3 \psi _3 \psi _2\},\ \{\widetilde{\phi }_3 \psi _3 \psi _4\}, \\ 
      & & \{\widetilde{\phi }_3 \psi _3 \psi _1\} \\ \cline{2-3}
      & \multirow{2}{*}{${\cal T}_8^5$} & \{\widetilde{\psi }_3 \widetilde{\phi }_5 \widetilde{\phi }_3\},\ \{\widetilde{\psi }_3 \widetilde{\phi }_2 \widetilde{\phi }_3\},\ \{\widetilde{\psi }_3^' \widetilde{\phi }_4 \widetilde{\phi }_3\},\  \{\widetilde{\psi }_3^' \widetilde{\phi }_1 \widetilde{\phi }_3\},\ \{\psi _5 \widetilde{\phi }_5 \widetilde{\phi }_3\},\ \{\psi _2 \widetilde{\phi }_2 \widetilde{\phi }_3\},\ \{\psi _4 \widetilde{\phi }_4 \widetilde{\phi }_3\}, \\
      & & \{\psi _1 \widetilde{\phi }_1 \widetilde{\phi }_3\} \\ \cline{2-3}
      & {\cal T}_8^6 & \{\widetilde{\phi }_3 \widetilde{\phi }_3\}\\ \cline{2-3}
      & {\cal T}_8^7 & \{\widetilde{\phi }_3 \psi _3 \phi _4\},\ \{\widetilde{\phi }_3 \psi _3 \phi _1\},\ \{\widetilde{\phi }_3 \widetilde{\psi }_4 \phi _4\},\ \{\widetilde{\phi }_3 \widetilde{\psi }_1 \phi _1\} \\ \cline{2-3}
      & \multirow{3}{*}{${\cal T}_8^8$} & \{\widetilde{\psi }_3^' \widetilde{\phi }_4 \widetilde{\psi }_3^'\},\ \{\widetilde{\psi }_3^' \widetilde{\phi }_1 \widetilde{\psi }_3^'\},\ \{\psi _4 \widetilde{\phi }_4 \widetilde{\psi }_3^'\},\ \{\psi _1 \widetilde{\phi }_1 \widetilde{\psi }_3^'\},\ \{\widetilde{\psi }_3 \widetilde{\phi }_5 \widetilde{\psi }_3\},\ \{\widetilde{\psi }_3 \widetilde{\phi }_2 \widetilde{\psi }_3\},\ \{\psi _5 \widetilde{\phi }_5 \widetilde{\psi }_3\}, \\ 
      & & \{\psi _2 \widetilde{\phi }_2 \widetilde{\psi }_3\},\ \{\widetilde{\psi }_3^' \widetilde{\phi }_4 \psi _4\},\ \{\widetilde{\psi }_3^' \widetilde{\phi }_1 \psi _1\},\ \{\psi _4 \widetilde{\phi }_4 \psi _4\},\ \{\psi _1 \widetilde{\phi }_1 \psi _1\},\  \{\widetilde{\psi }_3 \widetilde{\phi }_5 \psi _5\},\ \{\widetilde{\psi }_3 \widetilde{\phi }_2 \psi _2\}, \\
      & & \{\psi _5 \widetilde{\phi }_5 \psi _5\},\ \{\psi _2 \widetilde{\phi }_2 \psi _2\} \\ \hline \hline
      \multirow{13}{*}{\rotatebox{90}{\({\widetilde{\cal O}}_{8,D}^3 \equiv L^2 H^2 \bar\chi_R \chi_L\)}} 
      & {\cal T}_8^1  & \{\phi _5 \phi _5 \phi _1\},\ \{\phi _5 \widetilde{\phi }_3 \widetilde{\phi }_3^'\} \\ \cline{2-3}
      & {\cal T}_8^2  & \{\phi _1 \phi _3 \phi _5\},\ \{\phi _1 \phi _3 \phi _2\},\ \{\widetilde{\phi }_3^' \widetilde{\phi }_4 \widetilde{\phi }_3\},\ \{\widetilde{\phi }_3^' \widetilde{\phi }_1 \widetilde{\phi }_3\},\ \{\widetilde{\phi }_3\widetilde{\phi }_4\widetilde{\phi }_3^'\},\ \{\widetilde{\phi }_3\widetilde{\phi }_1\widetilde{\phi }_3^'\} \\ \cline{2-3}
      & \multirow{2}{*}{${\cal T}_8^3$}  & \{\psi _4\psi _4\phi _1\},\ \{\psi _1\psi _1\phi _1\},\ \{\psi _4\widetilde{\psi }_3^'\widetilde{\phi }_3^'\},\ \{\psi _1\widetilde{\psi }_3^'\widetilde{\phi }_3^'\},\ \{\psi _4\widetilde{\psi }_3\widetilde{\phi }_3\},\ \{\psi _1\widetilde{\psi }_3\widetilde{\phi }_3\},\ \{\widetilde{\psi }_3^'\widetilde{\psi }_3\phi _5\},  \\
      & & \{\widetilde{\psi }_3^'\widetilde{\psi }_3\phi _2\} \\ \cline{2-3}
      & \multirow{2}{*}{${\cal T}_8^4$} & \{\phi _1\psi _3\psi _4\},\ \{\phi _1\psi _3\psi _1\},\ \{\widetilde{\phi }_3^'\widetilde{\psi }_5^'\widetilde{\psi }_3^'\},\  \{\widetilde{\phi }_3^'\widetilde{\psi }_2^'\widetilde{\psi }_3^'\},\ \{\widetilde{\phi }_3\widetilde{\psi }_5\widetilde{\psi }_3\},\ \{\widetilde{\phi }_3\widetilde{\psi }_2\widetilde{\psi }_3\},\ \{\widetilde{\phi }_3^'\psi _3\psi _4\}, \\
      & &  \{\widetilde{\phi }_3^'\psi _3\psi _1\},\ \{\phi _5\widetilde{\psi }_5\widetilde{\psi }_3\},\ \{\phi _2\widetilde{\psi }_2\widetilde{\psi }_3\},\ \{\widetilde{\phi }_3\psi _3\psi _4\},\ \{\widetilde{\phi }_3\psi _3\psi _1\},\ \{\phi _5\widetilde{\psi }_5^'\widetilde{\psi }_3^'\},\ \{\phi _2\widetilde{\psi }_2^'\widetilde{\psi }_3^'\} \\  \cline{2-3}
      & \multirow{2}{*}{${\cal T}_8^5$} & \{\widetilde{\psi }_3\widetilde{\phi }_4\widetilde{\phi }_3\},\ \{\widetilde{\psi }_3\widetilde{\phi }_1\widetilde{\phi }_3\},\ \{\widetilde{\psi }_3^'\widetilde{\phi }_4\widetilde{\phi }_3^'\},\ \{\widetilde{\psi }_3^'\widetilde{\phi }_1\widetilde{\phi }_3^'\},\  \{\psi _4\phi _3\phi _1\},\ \{\psi _1\phi _3\phi _1\},\  \{\widetilde{\psi }_3\phi _3\phi _5\}, \\
      & & \{\widetilde{\psi }_3\phi _3\phi _2\},\ \{\psi _4\widetilde{\phi }_4\widetilde{\phi }_3^'\},\ \{\psi _1\widetilde{\phi }_1\widetilde{\phi }_3^'\},\ \{\widetilde{\psi }_3^'\phi _3\phi _5\},\ \{\widetilde{\psi }_3^'\phi _3\phi _2\},\ \{\psi _4\widetilde{\phi }_4\widetilde{\phi }_3\},\ \{\psi _1\widetilde{\phi }_1\widetilde{\phi }_3\}  \\ \cline{2-3}
      & {\cal T}_8^6 & \{\phi _5\phi _1\},\ \{\widetilde{\phi }_3\widetilde{\phi }_3^'\} \\ \cline{2-3}
      & {\cal T}_8^7 & \{\phi _1\psi _3\phi _5\},\ \{\widetilde{\phi }_3^'\psi _3\phi _5\},\ \{\widetilde{\phi }_3\psi _3\phi _5\},\ \{\widetilde{\phi }_3^'\widetilde{\psi }_5^'\phi _5\},\ \{\phi _5\widetilde{\psi }_5^'\phi _5\},\ \{\widetilde{\phi }_3\widetilde{\psi }_5\phi _5\},\ \{\phi _5\widetilde{\psi }_5\phi _5\}   \\ \cline{2-3}
      & \multirow{3}{*}{${\cal T}_8^8$} & \{\widetilde{\psi }_3\widetilde{\phi }_4\widetilde{\psi }_3^'\},\ \{\widetilde{\psi }_3\widetilde{\phi }_1\widetilde{\psi }_3^'\},\ \{\widetilde{\psi }_3^'\widetilde{\phi }_4\widetilde{\psi }_3\},\ \{\widetilde{\psi }_3^'\widetilde{\phi }_1\widetilde{\psi }_3\},\ \{\widetilde{\psi }_3\phi _3\psi _4\},\ \{\widetilde{\psi }_3\phi _3\psi _1\},\ \{\psi _4\widetilde{\phi }_4\widetilde{\psi }_3\}, \\
      & &  \{\psi _1\widetilde{\phi }_1\widetilde{\psi }_3\},\ \{\widetilde{\psi }_3^'\phi _3\psi _4\},\ \{\widetilde{\psi }_3^'\phi _3\psi _1\},\ \{\psi _4\widetilde{\phi }_4\widetilde{\psi }_3^'\},\ \{\psi _1\widetilde{\phi }_1\widetilde{\psi }_3^'\},\ \{\widetilde{\psi }_3\widetilde{\phi }_4\psi _4\},\ \{\widetilde{\psi }_3\widetilde{\phi }_1\psi _1\}, \\
      & &  \{\psi _4\widetilde{\phi }_4\psi _4\},\ \{\psi _1\widetilde{\phi }_1\psi _1\},\ \{\widetilde{\psi }_3^'\widetilde{\phi }_4\psi _4\},\ \{\widetilde{\psi }_3^'\widetilde{\phi }_1\psi _1\}\\ \hline \hline
      \multirow{8}{*}{\rotatebox{90}{\({\widetilde{\cal O}}_{8,D}^4 \equiv L^2 H^2 \bar\chi_L \chi_R\)}} 
      & {\cal T}_8^1  & \{\phi _5\phi _5\phi _1\} \\ \cline{2-3}
      & {\cal T}_8^2 & \{\phi _1\phi _3\phi _5\},\ \{\phi _1\phi _3\phi _2\} \\ \cline{2-3}
      & {\cal T}_8^3 & \{\psi _4\psi _4\phi _1\},\ \{\psi _1\psi _1\phi _1\},\ \{\widetilde{\psi }_3^'\widetilde{\psi }_3\phi _5\},\ \{\widetilde{\psi }_3^'\widetilde{\psi }_3\phi _2\} \\ \cline{2-3}
      & {\cal T}_8^4 & \{\phi _1\psi _3\psi _4\},\ \{\phi _1\psi _3\psi _1\},\ \{\phi _5\widetilde{\psi }_5\widetilde{\psi }_3\},\ \{\phi _2\widetilde{\psi }_2\widetilde{\psi }_3\},\ \{\phi _5\widetilde{\psi }_5^'\widetilde{\psi }_3^'\},\ \{\phi _2\widetilde{\psi }_2^'\widetilde{\psi }_3^'\} \\ \cline{2-3}
      & {\cal T}_8^5 & \{\psi _4\phi _3\phi _1\},\ \{\psi _1\phi _3\phi _1\},\ \{\widetilde{\psi }_3\phi _3\phi _5\},\ \{\widetilde{\psi }_3\phi _3\phi _2\},\ \{\widetilde{\psi }_3^'\phi _3\phi _5\},\ \{\widetilde{\psi }_3^'\phi _3\phi _2\} \\ \cline{2-3}
      & {\cal T}_8^6 & \{\phi _5\phi _1\} \\ \cline{2-3}
      & {\cal T}_8^7 & \{\phi _1\psi _3\phi _5\},\ \{\phi _5\widetilde{\psi }_5^'\phi _5\},\ \{\phi _5\widetilde{\psi }_5\phi _5\}\\ \cline{2-3}
      & {\cal T}_8^8 & \{\widetilde{\psi }_3\phi _3\psi _4\},\ \{\widetilde{\psi }_3\phi _3\psi _1\},\ \{\widetilde{\psi }_3^'\phi _3\psi _4\},\ \{\widetilde{\psi }_3^'\phi _3\psi _1\} \\ \hline
    \end{tabular}
    \caption{Tree-level UV completions for Dirac fermion DM. The second column lists the irreducible topologies, while the last column specifies representative sets of new heavy scalars ($\phi_i$, $\widetilde{\phi}_i$, $\widetilde{\phi}_i^\prime$) and fermions ($\psi_i$, $\widetilde{\psi}_i$, $\widetilde{\psi}_i^\prime$). The UV completions of the operators ${\cal \widetilde{O}}_{6,D}^2$ and ${\cal \widetilde{O}}_{8,D}^2$ are obtained by replacing $\widetilde{\phi}_i \leftrightarrow  \widetilde{\phi}_i^\prime$ and $\widetilde{\psi}_i \leftrightarrow  \widetilde{\psi}_i^\prime$. The operator ${\cal \widetilde{O}}_{8,D}^5$ has same UV completion as ${\cal \widetilde{O}}_{8,D}^3$. }
    \label{tab:DiracUV}
\end{table}
\clearpage
\noindent
introduced in renormalizable completions may themselves carry $U(1)$ charges $\pm q_\chi$. We adopt the notation $\{\widetilde{\phi}_i,\widetilde{\psi}_i\}$ for mediators with charge $+q_\chi$ and $\{\widetilde{\phi}'_i,\widetilde{\psi}'_i\}$ for those with charge $-q_\chi$. As in the Majorana scenario, we focus on the regime $m_\chi \lesssim v$, i.e.\ DM mass below the electroweak scale. 

The relevant DM-LEFT operators and their gauge-invariant DM-SMEFT embeddings are listed in Table~\ref{tab:operators}. A central difference compared to the Majorana case is that the additional global $U(1)$ symmetry forbids interactions containing an odd number of $\chi$ fields, and therefore eliminates entire classes of effective operators. As can be seen from Table~\ref{tab:operators}, the leading dimension-6 DM-LEFT interactions are in general generated only after EWSB from dimension-8 DM-SMEFT operators. The main exceptions are the direct matchings $\mathcal{O}_{6,D}^1 \to \widetilde{\mathcal{O}}_{6,D}^1$ and $\mathcal{O}_{6,D}^2 \to \widetilde{\mathcal{O}}_{6,D}^2$, for which dimension-6 gauge-invariant completions exist. Nevertheless, we also provide explicit UV completions for $\widetilde{\mathcal{O}}_{8,D}^1$ and $\widetilde{\mathcal{O}}_{8,D}^2$\footnote{These are higher-dimensional operators corresponding to $\mathcal{O}_{6,D}^{1,2}$ with $H^\dagger H$.}, since for certain mediator choices the first non-vanishing tree-level contributions arise only at dimension-8. The set of admissible renormalizable UV completions for the Dirac-fermion operator basis, together with their associated diagram topologies, is summarized in Table~\ref{tab:DiracUV}.

\subsection{Scalar DM}
\label{subsec:ScalarUV}
\begin{table}[!h]
\renewcommand{\arraystretch}{1.3}
\small
    \centering
    \begin{tabular}{|C|C|C|}
    \hline
    
     {\rm \bf Operator} & {\rm \bf Topology}  & {\rm \bf \hspace{4cm} New\ Particles}  \\
        \hline \hline
        \multirow{4}{*}{\rotatebox{90}{\({\widetilde{\cal O}}_{6,R}^1 \equiv L^2H^2\Phi \)}} 
      & {\cal T}_6^2  & \{\phi _3 \phi _5\},\ \{\phi _3 \phi _2\},\ \{\phi _5 \phi _5\} \\
        \cline{2-3}
      & {\cal T}_6^3  & \{\psi _4 \psi _3\},\ \{\psi _1 \psi _3\},\ \{\psi _4 \psi _4\}\\
        \cline{2-3}
      & {\cal T}_6^4  &  \{\phi _5 \psi _3\},\ \{\phi _3 \psi _4\},\ \{\phi _3 \psi _1\} \\
        \cline{2-3}
      &  {\cal T}_6^5 & \{\phi _5\} \\
        \hline \hline
      \multirow{10}{*}{\rotatebox{90}{\({\widetilde{\cal O}}_{7,R}^1 \equiv L^2H^2\Phi^2 \)}} 
      & {\cal T}_7^3  & \{\phi_5 \phi_5\},\ \{\phi_3 \phi_5\},\ \{\phi_3 \phi_2\},\ \{\phi_1 \phi_5\} \\ \cline{2-3}
      & {\cal T}_7^4  & \{\phi_5 \phi_1 \phi_5\},\ \{\phi_3 \phi_3 \phi_5\},\ \{\phi_3 \phi_3 \phi_2\} \\ \cline{2-3}
      & {\cal T}_7^5  & \{\phi_5 \phi_3 \phi_1\},\ \{\phi_2 \phi_3 \phi_1\},\ \{\phi_5 \phi_5 \phi_3\},\ \{\phi_2 \phi_2 \phi_3\},\ \{\phi_5 \phi_3 \phi_3\},\ \{\phi_2 \phi_3 \phi_3\},\ \{\phi_5 \phi_5 \phi_5\} \\ \cline{2-3}
      & {\cal T}_7^6  & \{\phi_5 \psi_3 \psi_3\},\ \{\phi_3 \psi_4 \psi_3\},\ \{\phi_3 \psi_1 \psi_3\},\ \{\phi_1 \psi_4 \psi_4\},\ \{\phi_1 \psi_1 \psi_1\} \\ \cline{2-3}
      & {\cal T}_7^7  & \{\psi_3 \psi_4 \psi_3\},\ \{\psi_3 \psi_1 \psi_3\},\ \{\psi_4 \psi_4 \psi_3\},\ \{\psi_1 \psi_1 \psi_3\},\ \{\psi_4 \psi_4 \psi_4\},\ \{\psi_1 \psi_1 \psi_1\} \\ \cline{2-3}
      & {\cal T}_7^8  & \{\psi_3 \phi_5 \phi_3\},\ \{\psi_3 \phi_2 \phi_3\},\ \{\psi_4 \phi_3 \phi_1\},\ \{\psi_1 \phi_3 \phi_1\},\ \{\psi_3 \phi_5 \phi_5\},\ \{\psi_4 \phi_3 \phi_3\},\ \{\psi_1 \phi_3 \phi_3\} \\ \cline{2-3}
      & {\cal T}_7^9  & \{\phi_1 \psi_3 \psi_4\},\ \{\phi_1 \psi_3 \psi_1\},\ \{\phi_3 \psi_4 \psi_3\},\ \{\phi_3 \psi_1 \psi_3\},\ \{\phi_3 \psi_4 \psi_4\},\ \{\phi_3 \psi_1 \psi_1\},\ \{\phi_5 \psi_3 \psi_3\} \\ \cline{2-3}
      & {\cal T}_7^{10}  &  \{\phi_5 \phi_3\},\ \{\phi_2 \phi_3\},\ \{\phi_5 \phi_5\} \\ \cline{2-3}
      & {\cal T}_7^{11}  & \{\psi_4 \phi_3\},\ \{\psi_1 \phi_3\},\ \{\psi_3 \phi_5\} \\ \cline{2-3}
      & {\cal T}_7^{12}  & \{\phi_1 \psi_3 \phi_5\},\ \{\phi_3 \psi_4 \phi_3\},\ \{\phi_3 \psi_1 \phi_3\} \\ \hline \hline
      \multirow{14}{*}{\rotatebox{90}{\({\widetilde{\cal O}}_{7,C}^1 \equiv L^2H^2\Phi^\dagger \Phi \)}} 
      & {\cal T}_7^3  & \{\phi_5 \phi_5\},\ \{\widetilde{\phi }_3 \phi_5\},\ \{\widetilde{\phi }_3 \phi_2\},\ \{\widetilde{\phi }_3^' \phi_5\},\ \{\widetilde{\phi }_3^' \phi_2\},\ \{\phi_1 \phi_5\} \\ \cline{2-3}
      & {\cal T}_7^4  & \{\phi_5 \phi_1 \phi_5\},\ \{\widetilde{\phi }_3 \widetilde{\phi }_3^' \phi_5\},\ \{\widetilde{\phi }_3 \widetilde{\phi }_3^' \phi_2\} \\ \cline{2-3}
      & \multirow{2}{*}{${\cal T}_7^5$}  & \{\phi_5 \phi_3 \phi_1\},\ \{\phi_2 \phi_3 \phi_1\},\ \{\phi_5 \widetilde{\phi }_5^' \widetilde{\phi }_3^'\},\ \{\phi_2 \widetilde{\phi }_2^' \widetilde{\phi }_3^'\},\ \{\phi_5 \widetilde{\phi }_5 \widetilde{\phi }_3\},\ \{\phi_2 \widetilde{\phi }_2 \widetilde{\phi }_3\},\ \{\phi_5 \phi_3 \widetilde{\phi }_3^'\},\ \\
      & &  \{\phi_2 \phi_3 \widetilde{\phi }_3^'\},\ \{\phi_5 \widetilde{\phi }_5 \phi_5\},\ \{\phi_5 \phi_3 \widetilde{\phi }_3\},\ \{\phi_2 \phi_3 \widetilde{\phi }_3\},\ \{\phi_5 \widetilde{\phi }_5^' \phi_5\} \\ \cline{2-3}
      & {\cal T}_7^6  & \{\phi_5 \widetilde{\psi }_3^' \widetilde{\psi }_3\},\ \{\widetilde{\phi }_3 \psi_4 \widetilde{\psi }_3\},\ \{\widetilde{\phi }_3 \psi_1 \widetilde{\psi }_3\},\ \{\widetilde{\phi }_3^' \psi_4 \widetilde{\psi }_3^'\},\ \{\widetilde{\phi }_3^' \psi_1 \widetilde{\psi }_3^'\},\ \{\phi_1 \psi_4 \psi_4\},\ \{\phi_1 \psi_1 \psi_1\} \\ \cline{2-3}
      & \multirow{2}{*}{${\cal T}_7^7$}  & \{\widetilde{\psi }_3 \widetilde{\psi }_4 \widetilde{\psi }_3^'\},\ \{\widetilde{\psi }_3 \widetilde{\psi }_1 \widetilde{\psi }_3^'\},\ \{\widetilde{\psi }_3^' \widetilde{\psi }_4 \widetilde{\psi }_3\},\ \{\widetilde{\psi }_3^' \widetilde{\psi }_1 \widetilde{\psi }_3\},\ \{\widetilde{\psi }_3 \psi_3 \psi_4\},\ \{\widetilde{\psi }_3 \psi_3 \psi_1\},\ \{\psi_4 \widetilde{\psi }_4 \widetilde{\psi }_3\},\  \\ 
      &  & \{\psi_1 \widetilde{\psi }_1 \widetilde{\psi }_3\},\ \{\widetilde{\psi }_3^' \psi_3 \psi_4\},\ \{\widetilde{\psi }_3^' \psi_3 \psi_1\},\ \{\psi_4 \widetilde{\psi }_4 \widetilde{\psi }_3^'\},\ \{\psi_1 \widetilde{\psi }_1 \widetilde{\psi }_3^'\},\ \{\psi_4 \widetilde{\psi }_4 \psi_4\},\ \{\psi_1 \widetilde{\psi }_1 \psi_1\}  \\ \cline{2-3}
      & \multirow{2}{*}{${\cal T}_7^8$}  & \{\widetilde{\psi }_3 \widetilde{\phi }_5 \widetilde{\phi }_3\},\ \{\widetilde{\psi }_3 \widetilde{\phi }_2 \widetilde{\phi }_3\},\ \{\widetilde{\psi }_3^' \widetilde{\phi }_5^' \widetilde{\phi }_3^'\},\ \{\widetilde{\psi }_3^' \widetilde{\phi }_2^' \widetilde{\phi }_3^'\},\ \{\psi_4 \phi_3 \phi_1\},\ \{\psi_1 \phi_3 \phi_1\},\ \{\widetilde{\psi }_3 \widetilde{\phi }_5 \phi_5\},\ \\
      & &  \{\psi_4 \phi_3 \widetilde{\phi }_3^'\},\ \{\psi_1 \phi_3 \widetilde{\phi }_3^'\},\ \{\widetilde{\psi }_3^' \widetilde{\phi }_5^' \phi_5\},\ \{\psi_4 \phi_3 \widetilde{\phi }_3\},\ \{\psi_1 \phi_3 \widetilde{\phi }_3\}  \\ \cline{2-3}
      & \multirow{2}{*}{${\cal T}_7^9$}  & \{\phi_1 \psi_3 \psi_4\},\ \{\phi_1 \psi_3 \psi_1\},\ \{\widetilde{\phi }_3^' \widetilde{\psi }_4 \widetilde{\psi }_3^'\},\ \{\widetilde{\phi }_3^' \widetilde{\psi }_1 \widetilde{\psi }_3^'\},\ \{\widetilde{\phi }_3 \widetilde{\psi }_4 \widetilde{\psi }_3\},\ \{\widetilde{\phi }_3 \widetilde{\psi }_1 \widetilde{\psi }_3\},\ \{\widetilde{\phi }_3^' \widetilde{\psi }_4 \psi_4\},\ \\
      & &  \{\widetilde{\phi }_3^' \widetilde{\psi }_1 \psi_1\},\ \{\phi_5 \psi_3 \widetilde{\psi }_3\},\ \{\widetilde{\phi }_3 \widetilde{\psi }_4 \psi_4\},\ \{\widetilde{\phi }_3 \widetilde{\psi }_1 \psi_1\},\ \{\phi_5 \psi_3 \widetilde{\psi }_3^'\}  \\ \cline{2-3}
      & {\cal T}_7^{10}  & \{\phi_5 \phi_3\},\ \{\phi_2 \phi_3\},\ \{\phi_5 \widetilde{\phi }_5^'\},\ \{\phi_5 \widetilde{\phi }_5\} \\ \cline{2-3}
      & {\cal T}_7^{11}  & \{\psi_4 \phi_3\},\ \{\psi_1 \phi_3\},\ \{\widetilde{\psi }_3^' \widetilde{\phi }_5^'\},\ \{\widetilde{\psi }_3 \widetilde{\phi }_5\} \\ \cline{2-3}
      & {\cal T}_7^{12}  & \{\phi_1 \psi_3 \phi_5\},\ \{\widetilde{\phi }_3^' \widetilde{\psi }_4 \widetilde{\phi }_3\},\ \{\widetilde{\phi }_3^' \widetilde{\psi }_1 \widetilde{\phi }_3\},\ \{\widetilde{\phi }_3 \widetilde{\psi }_4 \widetilde{\phi }_3^'\},\ \{\widetilde{\phi }_3 \widetilde{\psi }_1 \widetilde{\phi }_3^'\} \\ \hline 
    \end{tabular}
    \caption{Representative tree-level renormalizable UV completions for the scalar-DM-SMEFT operators $\widetilde{\cal O}_{6,R}^1$, $\widetilde{\cal O}_{7,R}^1$, and $\widetilde{\cal O}_{7,C}^1$. For each operator we list the relevant irreducible topologies ${\cal T}$ and example assignments of heavy scalar ($\phi_i$) and vector-like fermion ($\psi_i$) mediators using the notation of Table~\ref{tab:names}. Fields with tilde (prime) carry dark charge $+q_\Phi$ ($-q_\Phi$) in the complex-scalar case.}
    \label{tab:scalarUV}
\end{table}

We now turn to the case of scalar DM. We take the scalar DM $\Phi$ to be a singlet of SM with gauge quantum numbers $(1,1,0)$ under $SU(3)_c\times SU(2)_L\times U(1)_Y$. As in the case of Majorana and Dirac fermion DM, we focus on the regime $m_\Phi \lesssim v$ so that $\Phi$ remains a dynamical low-energy degree of freedom both above and below EWSB. We consider two
possibilities, which we denote by ``$R$'' and ``$C$'' in Table~\ref{tab:operators}:
(i) a self-conjugate (real) scalar, $\Phi=\Phi^\dagger$, and (ii) a complex
scalar carrying a conserved global $U(1)$ charge $q_\Phi$,
\begin{equation}
    \Phi \;\to\; e^{i q_\Phi \alpha}\,\Phi\,,
\end{equation}
so that operators must be neutral under this symmetry which forbids  operators of the type $\Phi$ and $\Phi^2$. In our operator-level discussion we remain agnostic about the specific DM-stabilizing symmetry (e.g.\ a $\mathbb{Z}_2$ parity), which may be specified once a concrete UV model is chosen.

The relevant DM-LEFT operators and their gauge-invariant DM-SMEFT embeddings are summarized in Table~\ref{tab:operators}. For scalar DM, the leading non-derivative DM-SMEFT structures that generate neutrino-DM interactions are, schematically,
\begin{align}
    \widetilde{\cal O}_{6,R}^1 &\equiv L^2 H^2 \Phi\,,
    \label{eq:scalarO6}\\[1mm]
    \widetilde{\cal O}_{7,R}^1 &\equiv L^2 H^2 \Phi^2\,,
    \qquad
    \widetilde{\cal O}_{7,C}^1 \equiv L^2 H^2 \Phi^\dagger\Phi\,,
    \label{eq:scalarO7}
\end{align}
where we have suppressed the Lorentz and gauge-index contractions for brevity. The operators  $\widetilde{\cal O}_{6,R}^1$ and $\widetilde{\cal O}_{7,R}^1$ involve $\Phi$ and $\Phi^2$, and therefore they  are only compatible with the complex-scalar scenario if the global $U(1)$ is absent or broken, whereas $\widetilde{\cal O}_{7,C}^1$ is automatically neutral under $U(1)$ and represents the leading interaction in the presence of a conserved dark charge. For the real-scalar case, both Eqs.~\eqref{eq:scalarO6} and \eqref{eq:scalarO7} are a priori allowed at the EFT level, with the caveat that additional stabilizing symmetries may forbid the operators that are odd in $\Phi$.

We construct renormalizable tree-level UV completions for the operators in
Eqs.~\eqref{eq:scalarO6}-\eqref{eq:scalarO7} using the topology classification
introduced above. The operator $\widetilde{\cal O}_{6,R}^1$ contains two external
fermions ($L^2$) and three external scalars ($H^2\Phi$), and thus is realized by
the $d=6$ topologies ${\cal T}_6^{2}$-${\cal T}_6^{5}$. The $d=7$ operators in
Eq.~\eqref{eq:scalarO7} contain two external fermions and four external scalars,
and therefore match onto the $d=7$ topologies ${\cal T}_7^{3}$-${\cal T}_7^{12}$.
In all cases, we assume that the internal lines correspond to heavy mediators that
can be integrated out. The mediator field content is built
from the scalar multiplets $\phi_i$ and vector-like fermion multiplets $\psi_i$
listed in Table~\ref{tab:names}. For the complex-scalar case, mediator fields
may carry dark charge $\pm q_\Phi$ in order to connect consistently to external
$\Phi$ and $\Phi^\dagger$ legs while preserving the global symmetry; we denote
fields with charge $+q_\Phi$ as $\{\widetilde{\phi}_i,\widetilde{\psi}_i\}$ and those with
charge $-q_\Phi$ as $\{\widetilde{\phi}'_i,\widetilde{\psi}'_i\}$, in analogy with the
Dirac-fermion-DM case  above.

Representative mediator assignments for each irreducible topology are collected
in Table~\ref{tab:scalarUV}.  
As before, we treat mediator
fields as distinguishable even when they share identical gauge quantum numbers;
this is convenient for bookkeeping, and it allows for the possibility that
multiple copies of certain fields are required, for instance when antisymmetric contractions would
otherwise force the amplitude to vanish.

\subsection{Vector DM}\label{subsec:VectorUV}
\begin{table}[]
\renewcommand{\arraystretch}{1.3}
\small
    \centering
    \begin{tabular}{|C|C|C|}
    \hline
      {\rm \bf Operator} & {\rm \bf Topology} & {\rm \bf \hspace{4cm} New\ Particles}  \\
        \hline\hline
     \multirow{2}{*}{\rotatebox{90}{\({\widetilde{\cal O}}_{6,V}^1  \)}}  & {\cal T}_6^3  & \{\psi_5 \psi_3\},\ \{\psi_2 \psi_3\},\ \{\psi_4 \psi_3\},\ \{\psi_1 \psi_3\},\ \{\psi_4 \psi_4\},\ \{\psi_1 \psi_1\},\ \{\psi_5 \psi_5\},\ \{\psi_2 \psi_2\} \\ \cline{2-3}
        & {\cal T}_6^4 & \{\phi_4 \psi_3\},\ \{\phi_1 \psi_3\} \\ \hline \hline 
        \multirow{11}{*}{\rotatebox{90}{\({\widetilde{\cal O}}_{7,V}^2 \equiv L^2 H^2 \chi_\mu \chi^\mu \)}} 
      & {\cal T}_7^3  & \{\phi_5 \phi_5\},\ \{\phi_1 \phi_5\} \\ \cline{2-3}
      & {\cal T}_7^4  & \{\phi_5 \phi_1 \phi_5\} \\ \cline{2-3}
      & {\cal T}_7^5  & \{\phi_5 \phi_3 \phi_1\},\ \{\phi_2 \phi_3 \phi_1\} \\ \cline{2-3}
      & {\cal T}_7^6  & \{\phi_5 \psi_3 \psi_3\},\ \{\phi_1 \psi_4 \psi_4\},\ \{\phi_1 \psi_1 \psi_1\} \\ \cline{2-3}
      & \multirow{2}{*}{${\cal T}_7^7$}  & \{\psi_3 \psi_4 \psi_3\},\ \{\psi_3 \psi_1 \psi_3\},\ \{\psi_3 \psi_3 \psi_4\},\ \{\psi_3 \psi_3 \psi_1\},\ \{\psi_4 \psi_4 \psi_3\},\ \{\psi_1 \psi_1 \psi_3\},\ \{\psi_4 \psi_4 \psi_4\},\ \\
      & & \{\psi_1 \psi_1 \psi_1\} \\ \cline{2-3}
      & {\cal T}_7^8  & \{\psi_4 \phi_3 \phi_1\},\ \{\psi_1 \phi_3 \phi_1\} \\ \cline{2-3}
      & {\cal T}_7^9  & \{\phi_1 \psi_3 \psi_4\},\ \{\phi_1 \psi_3 \psi_1\},\ \{\phi_5 \psi_3 \psi_3\} \\ \cline{2-3}
      & {\cal T}_7^{10}  & \{\phi_5 \phi_3\},\ \{\phi_2 \phi_3\} \\ \cline{2-3}
      & {\cal T}_7^{11}  & \{\psi_4 \phi_3\},\ \{\psi_1 \phi_3\} \\ \cline{2-3}
      & {\cal T}_7^{12}  & \{\phi_1 \psi_3 \phi_5\} \\ \hline 
    \end{tabular}
    \caption{Representative tree-level UV completions for vector DM-SMEFT operators. For each operator we list the contributing irreducible topology ${\cal T}$ and example heavy-mediator assignments in terms of $\phi_i$ and $\psi_i$ (Table~\ref{tab:names}). The upper (lower) block corresponds to the single-vector operator $\widetilde{\cal O}_{6V}^1$ (the even-vector operator $\widetilde{\cal O}_{7V}^2$).}

    \label{tab:placeholder}
\end{table}

We finally consider the case of vector-boson DM, denoted by $\chi_\mu$. We take $\chi_\mu$ to be a SM-singlet with gauge quantum numbers $(1,1,0)$ under $SU(3)_c\times SU(2)_L\times U(1)_Y$, and we assume $m_\chi\lesssim v$ so that $\chi_\mu$ remains a dynamical degree of freedom in both the DM-SMEFT and DM-LEFT descriptions. At the effective level, a convenient starting point is a massive  vector field with field strength $\chi_{\mu\nu}\equiv \partial_\mu\chi_\nu-\partial_\nu\chi_\mu$ and Lagrangian
\begin{equation}
  \mathcal{L}\supset -\frac14\,\chi_{\mu\nu}\chi^{\mu\nu}
+\frac12\,m_\chi^2\,\chi_\mu\chi^\mu \, ,
\end{equation}
while the UV interpretation can be that $\chi_\mu$ originates as a gauge boson of a broken dark gauge symmetry. We remain agnostic about the precise stabilizing symmetry (e.g.\ a dark parity or a remnant discrete symmetry after spontaneous breaking).

The relevant DM-LEFT operators involving neutrinos and their gauge-invariant DM-SMEFT embeddings are summarized in Table~\ref{tab:operators}. A notable feature of vector DM is that gauge invariance already permits a renormalizable dimension-4 interaction,
\begin{equation}
  \widetilde{\mathcal O}_{4V}^1 \equiv (\bar L\gamma_\mu L)\,\chi^\mu \, ,
\end{equation}
which matches onto ${\mathcal O}_{4V}^1$ after EWSB. However, if $\chi_\mu$ is to be a viable long-lived dark-matter candidate, such a linear coupling must either be forbidden by the stabilizing symmetry (for instance by a $\mathbb{Z}_2$ under which $\chi_\mu$ is odd) or be extremely suppressed to avoid rapid decays $\chi\to \nu\bar\nu$. This criterion motivates our particular attention to the leading operators containing an even number of $\chi_\mu$ fields, which can naturally be compatible with a dark parity.

In this subsection, we therefore provide explicit tree-level renormalizable UV completions for the lowest-dimensional non-derivative vector-DM operators of interest, using the topology classification discussed above. The operator with a single vector boson,
\begin{equation}
  \widetilde{\mathcal O}_{6V}^1 \equiv
  \big(H^\dagger\,\bar L\gamma_\mu\,\vec L\!\cdot\!\vec H\big)\,\chi^\mu\,,
\end{equation}
contains two external fermions and three external bosons ($H$, $H^\dagger$ and $\chi_\mu$), and is realized by the corresponding $d=6$ topologies (${\cal T}_6^3$ and ${\cal T}_6^4$ here). The leading even-$\chi_\mu$ structure relevant for neutrino-DM scattering,
\begin{equation}
  \widetilde{\mathcal O}_{7V}^2 \equiv
  \big(\vec L\!\cdot\!\vec H\ \vec L\!\cdot\!\vec H\big)\,\chi_\mu\chi^\mu\,,
\end{equation}
has two external fermions and four external bosons (two Higgs doublets and two vector fields) and is therefore generated by the $d=7$ topologies ${\cal T}_7^{3}$-${\cal T}_7^{12}$. As before, we suppress Lorentz and $SU(2)_L$ contractions and the automated UV-completion procedure accounts for all allowed attachments of external fields and all inequivalent contractions for a given topology.

Renormalizable couplings of $\chi_\mu$ to the mediator sector can then arise through conserved currents, e.g.\ $\chi_\mu\,\bar\psi\gamma^\mu\psi$ and/or $\chi_\mu\,\phi^\dagger i\overleftrightarrow{\partial^\mu}\phi$, as would be expected if $\chi_\mu$ originates from a dark gauge symmetry under which the mediators are charged. Representative mediator assignments for each irreducible topology are collected in Table~\ref{tab:placeholder}. Curly braces indicate the set of heavy fields propagating internally for a given topology; as in the other cases, we treat mediators as distinguishable even when they share the same SM gauge quantum numbers, since repeated representations may be required to avoid vanishing amplitudes from antisymmetric contractions.

\section{Model 1: Pseudo-Dirac DM} \label{sec:model1}
Having systematically classified the UV completions of each operator by their irreducible topologies, we now turn to \emph{minimal model} realizations that can generate sizable neutrino-DM interactions. We restrict our attention to renormalizable extensions of the SM\,+\,DM field content by one or two new states (scalar or fermion), with gauge quantum numbers compatible with Table~\ref{tab:names}  and an implicit DM–stabilizing symmetry. For each model, we specify the mediator content, the interaction Lagrangian, the matching onto the effective operator(s), and the parametric size of the induced coupling. We focus on those realizations that, in principle, can produce comparatively large DM–neutrino interactions while remaining phenomenologically viable.

We begin with the minimal setup that introduces an inert second Higgs doublet $\phi_3 \equiv H_2 \sim (1,2,1/2)$~\cite{Deshpande:1977rw,Barbieri:2006dq}, in addition to the singlet Majorana DM field $\chi$~\cite{Tao:1996vb, Ma:2006km}, which is the well-known {\it scotogenic model}. A stabilizing discrete $\mathbb{Z}_2$ symmetry is imposed under which $\{\chi,H_2\}\to -\{\chi,H_2\}$ and all the SM fields are taken to be $\mathbb{Z}_2$–even. This forbids the bilinear term $H_1^\dagger H_2$, enforces $\langle H_2\rangle=0$, and prevents the mixing between $H_1$ and $H_2$. The interactions relevant for generating the effective operators in Table~\ref{tab:operators} arise from the Yukawa and scalar potential terms and can be written as
\begin{align}
\label{eq:L_MajH2}
\mathcal{L}_{\text{int}} \;\supset\;
  \left[  y_\chi^i\ {L_i}^T C \widetilde{H}_2^* \chi_L + {\rm H.c.} \right]
 - V(H_1,H_2) ,
\end{align}
where $L_i$ denotes the SM lepton doublet of $i$-th family and $H \equiv H_1$ is the SM Higgs doublet. The scalar potential term $V(H_1, H_2)$ in the $\mathbb{Z}_2$-symmetric limit is given by
\begin{align}
V(H_1,H_2) \;=\;&
m_1^2\, H_1^\dagger H_1 + m_2^2\, H_2^\dagger H_2
+ \frac{\lambda_1}{2} (H_1^\dagger H_1)^2 + \frac{\lambda_2}{2} (H_2^\dagger H_2)^2
\nonumber\\
&+ \lambda_3\, H_1^\dagger H_1\, H_2^\dagger H_2
+ \lambda_4\, H_1^\dagger H_2\, H_2^\dagger H_1
- \left[\frac{\lambda_5}{2}\, (H_1^\dagger H_2)^2 + \text{H.c.}\right].
\label{eq:V_IDM}
\end{align}
EWSB occurs when the neutral component of $H_1$ gets a VEV, $\langle H_1\rangle=(0,v/\sqrt{2})^T$.  
In component form, $H_1$ and $H_2$  read as
\begin{equation}
H_1 \;=\; \begin{pmatrix} G^+ \\ \frac{1}{\sqrt{2}}(h + i\ G^0 + v)\end{pmatrix}, \hspace{5mm} H_2 \;=\; \begin{pmatrix} H_2^+ \\ \frac{1}{\sqrt{2}}(H_2^0 + i A)\end{pmatrix},
\end{equation}
where $\{G^+, G^0\}$ are the Goldstone modes and $h$ is the physical SM Higgs field. 
After EWSB, the mass of the SM Higgs is given by $m_h^2 = \lambda_1 v^2$ and the mass eigenvalues for the inert-doublet fields are given by
\begin{align}
m_{H_2^\pm}^2 &= m_2^2 + \frac{\lambda_3 v^2}{2}, \quad
m_{H_2^0}^2 \;=\; m_2^2 + \frac{(\lambda_3+\lambda_4-\lambda_5) v^2}{2}, \quad
m_{A}^2 \;=\; m_{H_2^0}^2 + \lambda_5 v^2.
\label{eq:IDM_masses}
\end{align}
In the limit of $\lambda_5 \to 0$, the neutral inert CP-even scalar $H_2^0$ and CP-odd scalar $A$ become degenerate in mass. The scalar quartic couplings of Eq.~\eqref{eq:V_IDM} have to comply with the vacuum stability conditions~\cite{Deshpande:1977rw,Kannike:2012pe}
\begin{equation}
    \lambda_1 \geq 0, \hspace{3mm} \lambda_2 \geq 0, \hspace{3mm} \lambda_3 + \sqrt{\lambda_1 \lambda_2} \geq 0, \hspace{3mm}  \lambda_3 + \lambda_4 - \lambda_5 + \sqrt{\lambda_1 \lambda_2} \geq 0 \, ,
\end{equation}
and the perturbativity condition implies that the $\lambda_5 \leq 8 \pi$~\cite{Kanemura:1999xf,Akeroyd:2000wc}. Note that both CP-even $H_2^0$ and CP-odd $A$ cannot be arbitrarily split from the charged component 
$H_2^+$ due to electroweak precision constraints. In particular, the electroweak $T$-parameter~\cite{Peskin:1990zt,Peskin:1991sw} constrains the mass splitting between different components  of the inert doublet: 
\begin{equation}
    \Delta T \simeq \frac{1}{12 \pi^2 \alpha_{\rm em} v^2} (m_{H_2^+} - m_A) (m_{H_2^+} - m_{H_2^0}) \, ,
    \label{eq:Tparam}
\end{equation}
where $\alpha_{\rm em}$ is the electromagnetic fine structure constant. We are interested in maximizing the allowed neutrino-DM interaction, we choose to work with a light CP-even neutral scalar mediator limit $m_{H_2^0}\lesssim v$. Since the charged component cannot be very light (see e.g., Refs.~\cite{
Ahriche:2018ger, Babu:2019mfe, Herms:2023cyy}), we go to the limit $m_A \simeq m_{H_2^+}\gg m_{H_2^0}$ to ensure that the contribution to $\Delta T$ remains within the experimentally allowed range $\Delta T = 0.01 \pm 0.12$~\cite{ParticleDataGroup:2024cfk}.

\begin{figure}[!t]
    \centering
    \begin{subfigure}[b]{0.2\linewidth}
        \includegraphics[width=\linewidth]{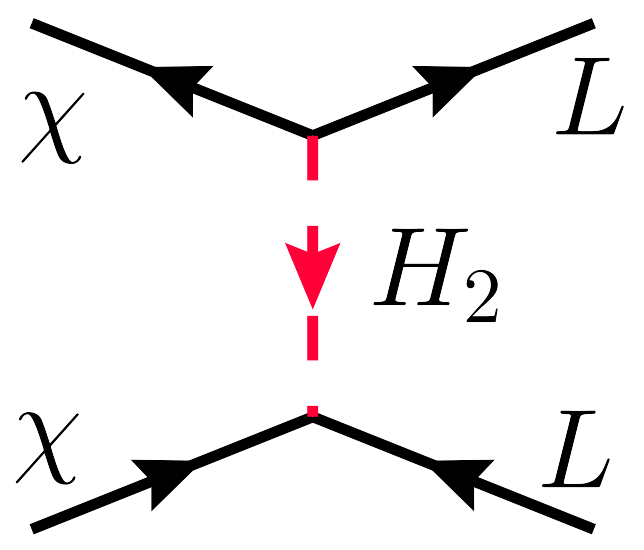}
        \caption{}
    \end{subfigure}
    \hspace{1.5cm}
    \begin{subfigure}[b]{0.2\linewidth}
        \includegraphics[width=\linewidth]{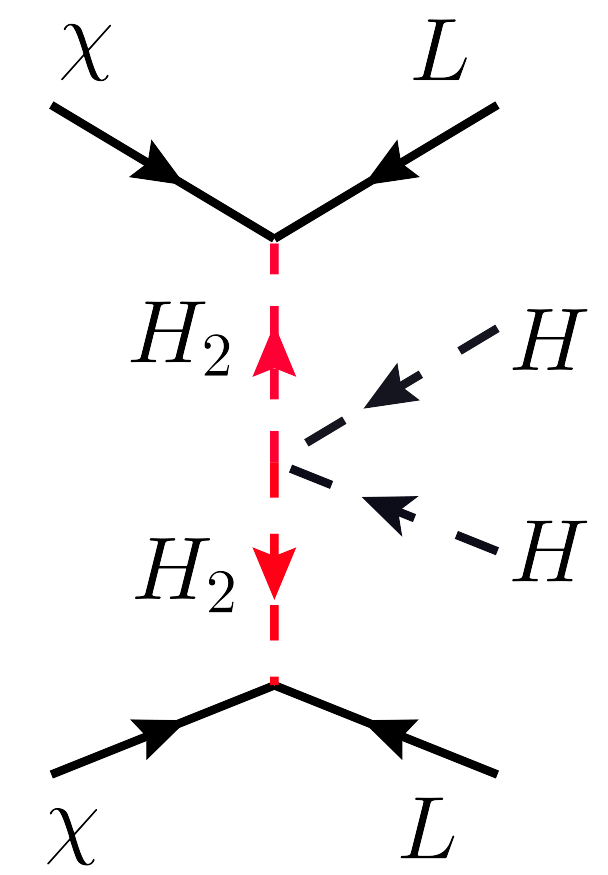}
        \caption{}
    \end{subfigure}
    \caption{Feynman diagrams illustrating the UV completions of effective operators arising in Model~1 of Section~\ref{sec:model1}, which introduces an additional Higgs doublet $\phi_3 \equiv H_2$. 
    (a), and (b) respectively represent the UV completion of 
    ${\cal \widetilde{O}}_{6,M}^3 \equiv (\bar{L}\gamma_\mu L)(\bar{\chi_L}\gamma^\mu\chi_L)$ 
    and ${\cal \widetilde{O}}_{8,M}^1 \equiv (\vec L\!\cdot\!\vec H\,\vec L\!\cdot\!\vec H)(\chi_L\chi_L)$.}
    \label{fig:tree_model1}
\end{figure}
The Yukawa coupling $y_\chi$ and the quartic couplings between $H_1$ and $H_2$ generate the leading effective operators once the heavy scalar inert doublet $H_2$ is integrated out. At tree level, this matching induces dimension-6 and dimension-8 operator 
\begin{equation}
    {\cal \widetilde{O}}_{6,M}^3 \equiv (\bar{L}\gamma_\mu L)(\bar{\chi_L}\gamma^\mu\chi_L) \, , \hspace{5mm} {\cal \widetilde{O}}_{8,M}^1 \equiv (\vec L\!\cdot\!\vec H\,\vec L\!\cdot\!\vec H)(\chi_L\chi_L),\
\end{equation}
They  arise through the tree-level diagrams shown in 
Fig.~\ref{fig:tree_model1}. The corresponding Wilson coefficients obtained by matching at $\mu \simeq m_{H_2}$ (which generically stands for the mass of the relevant $H_2$ doublet field component) are
\begin{equation}
({\widetilde C}_{6,M}^3)^{ij} = 
-\frac{y_\chi^i y_\chi^{j*}}{2 m_{H_2}^2}\, , \hspace{5mm}
({\widetilde C}_{8,M}^1)^{ij} = 
-\frac{y_\chi^i y_\chi^j \,\lambda_5}{4 m_{H_2}^4}\, . 
\end{equation}
Here ${\widetilde C}_{6,M}^3$ and ${\widetilde C}_{8,M}^1$ denote the coefficients of the operators $\widetilde{\mathcal{O}}^{\,3}_{6,M}$ and $\widetilde{\mathcal{O}}^{\,1}_{8,M}$, respectively. After EWSB, the operator ${\cal \widetilde{O}}_{6,M}^3$ matches directly on to the LEFT operator involving neutrinos
\begin{equation}
    (C_{6,M}^3)^{ij} = 2\ ({\widetilde C}_{6,M}^3)^{ij} \, .
\end{equation}
The dimension-8 operators $\widetilde{\mathcal{O}}^{\,1}_{8,M}$ give rise to the scalar LEFT operators after inserting two Higgs VEVs, each contributing a factor of $v/\sqrt{2}$. Their Wilson coefficients are therefore
\begin{equation}
    C_{6,M}^{1} = \frac{v^2}{2}\ {\widetilde C}_{8,M}^1\, \, .
    \label{eq:wilsonC6M}
\end{equation}
We note that these expressions are valid when the components of $H_2$ have a common heavy mass $m_{H_2}$. 

When $\lambda_5 \to 0$, the CP-even and CP-odd neutral components of $H_2$ become nearly degenerate. In this limit, the electroweak $T$-parameter constrains the mass splitting between the charged and neutral scalars to be at most  $40~\text{GeV}$. The charged state $H_2^+$ then decays almost exclusively via $H_2^+ \to \ell^+ \chi$, producing a signature that closely resembles left-handed slepton pair production in supersymmetric scenarios. Existing LHC Run-2 searches from ATLAS~\cite{ATLAS:2019lff} and CMS~\cite{CMS:2020bfa} set a lower bound of $m_{H_2^+} \gtrsim 700~\text{GeV}$ at 95\%~C.L. for couplings to electron and muon flavors (350 GeV for coupling to tau flavor~\cite{CMS:2022syk, ATLAS:2024fub}), which directly applies to the region of interest here with  $m_\chi \ll m_{H_2^+}$. However, it is important to point out that the mass range of $80-125$ GeV has not been fully excluded for a leptophilic charged scalar, depending on the branching fraction of the charged scalar into different lepton flavors~\cite{Babu:2019mfe,Das:2025oww}; therefore, the $H_2^+$ component in our case can still be as light as ${\cal O}(100)~{\rm GeV}$.

\begin{figure}
    \centering
    \includegraphics[width=0.4\linewidth]{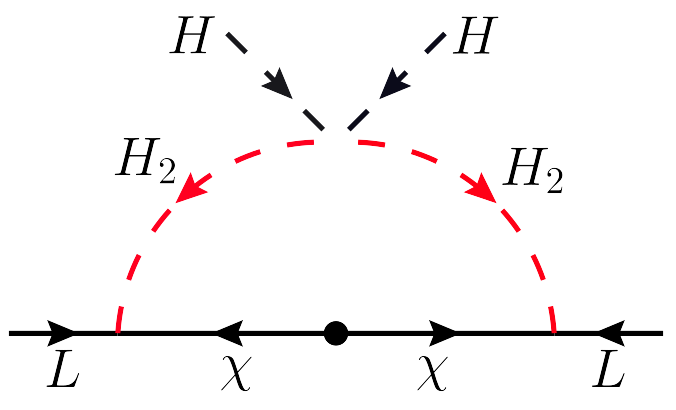}
    \caption{One-loop neutrino mass in the scotogenic model through the effective dimension-5 operator $LLHH$. }
    \label{fig:scoto-nu}
\end{figure}

Furthermore, in the $\lambda_5 \to 0$ limit, lepton number symmetry is restored and the one-loop radiative neutrino mass vanishes (see Fig.~\ref{fig:scoto-nu}). As a result, the dominant contribution to the low-energy effective interaction arises from the coefficient $C_{6,M}^3$, while the dimension-6 term $C_{6,M}^1$ is suppressed primarily because $\lambda_5 \to 0$. However, since the inert scalar fields are necessarily heavier than the electroweak scale, this scenario prevents any significant enhancement of the DM-neutrino effective coupling, even though $C_{6,M}^3$ is parametrically dominant.

In contrast, for larger values of $\lambda_5 \sim {\cal O}(1)$, the mass splitting between the neutral CP-even scalar $H_2^0$ and the pseudoscalar $A$ becomes significant, leading to a clear hierarchy $m_{H_2^0} \ll m_{H_2^+}\simeq m_A$, which is crucial for us to achieve large neutrino-DM interactions via light $H_2^0$, while the corresponding DM-charged lepton interactions mediated by the same Yukawa couplings are suppressed by the heavy $H_2^+$. The condition $m_{H_2^+} \simeq m_A$ is imposed by constraints from the electroweak $T$-parameter [cf.~Eq.~\eqref{eq:Tparam}]. In this situation, the effective theory must be constructed in the broken phase: integrating out $H_2^+$, $A$, and $H_2^0$. The effective DM–LEFT Wilson coefficients below $m_{H_2^0}$ are
\begin{align}
(C_{6,M}^3)^{ij} &= -\frac{y_\chi^i y_\chi^{j*}}{2}
\left(\frac{1}{m_{H_2^0}^2} + \frac{1}{m_A^2}\right),
\label{eq:Lcons}
\\
(C_{6,M}^{1})^{ij} &= -\frac{y_\chi^i y_\chi^{j} }{8}
\left(\frac{1}{m_{H_2^0}^2} - \frac{1}{m_A^2}\right) = -\frac{y_\chi^i y_\chi^{j} }{8} \frac{\lambda_5 v^2}{m_{H_2^0}^2 m_A^2},
\label{eq:Lviol}
\end{align}
where Eq.~\eqref{eq:Lcons} and Eq.~\eqref{eq:Lviol} are lepton number conserving and lepton number violating effective interactions. Note that Eq.~\eqref{eq:Lviol} vanishes in the limit $m_{H_2^0} = m_A$, implying lepton number conservation through $\lambda_5 = 0$. For comparison, setting $m_{H_2^0} = 50$ MeV and Yukawa coupling $y_\chi^{33} = 0.01$ (a value consistent with the observed relic abundance, see Fig.~\ref{fig:money_model1}) leads to a sizable enhancement of neutrino-DM interaction, corresponding to $|C_{6,M}^{3}| \simeq 1.7\times 10^{3}\ G_F$ and $|C_{6,M}^{1}| \simeq 4.3\times 10^{2}\ G_F$.

\subsection{Radiative neutrino mass}
So far, we have discussed how to make the scalar field $H_2^0$ light in order to enhance the neutrino-DM interactions in the scotogenic model. Now let us address how to make (some of) the Yukawa couplings $y_\chi^i$ large. For one generation of $\chi$, the strongest limit on $y_\chi$ comes from the neutrino mass. Since lepton number is broken, neutrinos acquire mass via the scotogenic one-loop diagram shown in  Fig.~\ref{fig:scoto-nu}, which forces the Yukawa coupling  to be small. With three generations of neutrinos, the neutrino mass matrix  is given by 
\begin{equation}
(m_\nu)_{ij}= y_\chi^i\,y_\chi^j\, {\cal F} ,
\end{equation}
with the loop factor~\cite{Ma:2006km} 
\begin{equation}
{\cal F}=\frac{m_\chi}{32\pi^2}\left[
\frac{m_{H_2^0}^2}{m_{H_2^0}^2-m_\chi^2}\ln\!\frac{m_{H_2^0}^2}{m_\chi^2}
-\frac{m_A^2}{m_A^2-m_\chi^2}\ln\!\frac{m_A^2}{m_\chi^2}
\right].
\end{equation}
With one $\chi$, the neutrino mass matrix is rank-1 and, in the hierarchical limit $m_\chi \ll m_{H_2^0} \ll m_A$, the loop factor simplifies to
\begin{equation}
{\cal F} \simeq \frac{m_\chi}{32\pi^2}\,\ln\!\frac{m_{H_2^0}^2}{m_A^2}\,,
\end{equation}
so the Yukawa coupling required to generate the observed neutrino mass is
\begin{equation}
|y|
\;\lesssim\;
4\times 10^{-3}\;
\left(\frac{m_\nu}{0.05~\text{eV}}\right)^{\!1/2}
\left(\frac{\text{MeV}}{m_\chi}\right)^{\!1/2}
 \left[  \ln \left(\frac{m_A^2}{m_{H_2^0}^2} \right) \right]^{-1/2}
. 
\end{equation}
However, a single generation of $\chi$ cannot reproduce the observed pattern of neutrino oscillations. To account for the measured mass splittings and mixing angles, one must either introduce an additional mechanism for neutrino mass generation or extend the model to include multiple generations of fermionic fields, as in the standard scotogenic scenario. The latter option is particularly appealing because, with an appropriate flavor structure~\cite{Fan:2023cml}, it allows Yukawa couplings of order unity while still naturally explaining the smallness of neutrino masses.
With 3 generations of $\chi$ fields (usually denoted as $N_{1,2,3}$ in the scotogenic literature), the neutrino mass matrix takes the form
\begin{equation}
    m_\nu = y_\chi {\cal F} y_\chi^T \, ,
\end{equation}
where both $y_\chi$ and ${\cal F} \sim m_\chi$ are now $3\times 3$ matrices. With the form
\begin{equation}
    y_\chi = \begin{pmatrix}
    \epsilon_1 & \epsilon_2 & y_\chi^{13} \\
    \epsilon_1 & \epsilon_2 & y_\chi^{23} \\
     \epsilon_1 & \epsilon_2 & y_\chi^{33}
\end{pmatrix} \, , \hspace{5mm} 
m_\chi = \begin{pmatrix}
    M_1 & \widetilde{\epsilon} & \widetilde{\epsilon} \\
    \widetilde{\epsilon} & \widetilde{\epsilon} & M_2 \\
     \widetilde{\epsilon} & M_2 & \widetilde{\epsilon}
\end{pmatrix} \, ,
\label{eq:Yukmat}
\end{equation}
one can see that in the limit $\epsilon_{1,2},\widetilde\epsilon\to0$, the active neutrinos remain massless, while the singlet sector contains one Majorana state with mass $|M_1|$ and a pseudo-Dirac pair with mass $|M_2|$ (equivalently two Majorana eigenvalues $\pm M_2$). The form in Eq.~\eqref{eq:Yukmat} is similar to the texture used in Ref.~\cite{Kersten:2007vk} to enhance the light-heavy neutrino mixing. Switching on the small parameters $\epsilon_{1,2}$ and $\widetilde{\epsilon}$ breaks the exact cancellation and generates a nonzero neutrino mass. At leading order, the neutrino mass can be written schematically as
\begin{equation}
(m_\nu)_{ij}\;\sim\;
\frac{1}{32\pi^2}\,
\ln\!\left(\frac{m_A^2}{m_{H_2^0}^2}\right)
\Big[
\widetilde{\epsilon}\,y_\chi^{i3}y_\chi^{j3}
+\epsilon_2 M_2 (y_\chi^{i3} + y_\chi^{j3})
+ M_1\epsilon_1^2
\Big].
\end{equation}
In this scenario,  we can take large $y_\chi^{i3}$ even  up to the perturbative limit while being consistent with the neutrino mass constraint. For instance, taking $y_\chi^{i3}\sim 1$ with $m_A = 1$ TeV and $m_{H_2^0} = 10$ MeV, we find the size of the lepton-number breaking parameters required to generate the observed neutrino mass scale $m_\nu \sim 0.1$ eV to be $\widetilde{\epsilon} \sim {\cal O} (10^{-9})$ GeV and $\epsilon_2 \sim {\cal O} (10^{-7})$. 

\subsection{Lepton flavor violation}
As mentioned in the previous section, one could allow $\mathcal{O}(1)$ entries for $y_\chi^{i3}$ without altering the prediction for light-neutrino masses, i.e., in the limit $\epsilon_i,\widetilde{\epsilon}\to 0$ active neutrinos remain massless. However, such off-diagonal Yukawa couplings induce charged-lepton flavor violation (cLFV) through the radiative decays $\ell_i\to\ell_j\gamma$ at one loop, with the $(H_2^\pm,\chi_a)$ states in the loop~\cite{Lavoura:2003xp}. Using  the small-mass-ratio limit $m_\chi^2/m_{H_2^+}^2\ll 1$, the experimental upper limits on the branching ratios of the radiative decays  $\mu\to e\gamma$~\cite{MEGII:2025gzr}, $\tau\to e\gamma$~\cite{BaBar:2009hkt}, and $\tau\to\mu\gamma$~\cite{Belle:2021ysv} imply the following constraints on products of Yukawa couplings mediated by the charged scalar $H_2^+$:
\begin{align}
    |y_\chi^{13}\,y_\chi^{23}|
    &< 2.59\times 10^{-5}\,
    \left(\frac{m_{H_2^+}}{100~{\rm GeV}}\right)^{2}, \\[4pt]
    |y_\chi^{13}\,y_\chi^{33}|
    &< 2.88\times 10^{-2}\,
    \left(\frac{m_{H_2^+}}{100~{\rm GeV}}\right)^{2}, \\[4pt]
    |y_\chi^{23}\,y_\chi^{33}|
    &< 3.33\times 10^{-2}\,
    \left(\frac{m_{H_2^+}}{100~{\rm GeV}}\right)^{2}.
\end{align}

\subsection{DM relic abundance}
As discussed above, the two light singlets $\chi_2$ and $\chi_3$ form a pseudo-Dirac pair.\footnote{We work in the basis $X_L=\chi_2$, $X_R=\chi_3^{\,c}$ so that $X=(X_L,X_R)^T$ is a Dirac fermion with a lepton number preserving mass $M$ and small lepton number violating parameters $\mu_{L,R}\ll M$ that generate the tiny splitting $\delta m\sim|\mu_L+\mu_R|$. In the freeze–out epoch we assume $\delta m\ll T_f$, so the two Majorana components remain in chemical equilibrium and behave as a single Dirac species.} 
In the case of specific structure given in Eq.~\eqref{eq:Yukmat}, the Yukawa interaction is carried by the \emph{right–handed} component of $\chi$:
\begin{equation}
\label{eq:Dirac-int}
\mathcal L \;\supset\;
\frac{y_\chi^{i 3}}{\sqrt{2}}\ \nu_{Li}^{T} C\, H_2^0\, \chi
\;-\;
y_\chi^{i 3}\ \ell_{Li}^{T} C\, H_2^+\, \chi
\;+\; \text{H.c.}\ ,
\end{equation}
As discussed below Eq.~\eqref{eq:IDM_masses}, we consider the split spectrum $m_{H_2^0}\ll m_{H_2^+}\simeq m_A$. In this regime, the dominant annihilation channel at DM freeze-out is
\begin{equation}
\chi\,\overline{\chi} \ \to\ \nu_i\,\overline{\nu}_j
\qquad\text{via $t$-channel exchange of $H_2^0$}\,,
\label{eq:424}
\end{equation}
whereas channels involving $H_2^\pm$ and $A$ are suppressed by additional powers of $m_{H_2^0}^2/m_{H_2^\pm}^2$ and $m_{H_2^0}^2/m_A^2$ respectively. Because the initial state is (pseudo)Dirac and the final-state leptons are (nearly) massless, annihilation is \emph{$s$-wave} dominated. The thermally averaged cross section can be written as~\cite{Gondolo:1990dk,Berlin:2014tja}
\begin{equation}
\langle \sigma v \rangle \;\simeq\; \frac{1}{128\pi}\;
\frac{|y_{\chi}^{i 3}|^4\, m_\chi^{2}}{\big(m_{H_2^0}^2 + m_\chi^2\big)^{2}} \,.
\label{eq:a-Dirac}
\end{equation}

Neglecting coannihilations with the heavier inert scalars and possible narrow resonances.\footnote{If $m_{H_2^0}-m_\chi\lesssim T_f$, coannihilations $\chi\,H_2^0\to \text{SM}$ and $H_2^0 H_2^0\to \text{SM}$ should be included. Our results remain unchanged as long as $m_{H_2^0}-m_\chi\gtrsim \mathcal{O}(10\%)\,m_\chi$.} the standard $s$–wave solution yields
\begin{equation}
\Omega_\chi h^2 \;\simeq\;
\frac{1.07\times 10^{9}\ \text{GeV}^{-1}}{M_{\rm Pl}}\,
\frac{x_f}{\sqrt{g_\star (T_f)}}\;\frac{1}{\langle \sigma v \rangle} \,,
\label{eq:Omega-swave}
\end{equation}
where $x_f \equiv m_\chi/T_f$ is the value at freeze-out temperature $T_f$, which is found by iteration~\cite{Steigman:2012nb, Berlin:2014tja} and  $g_\star(T_f)$ denotes the temperature-dependent effective number of relativistic degrees of freedom at freeze-out, for which we use the data from Ref.~\cite{Saikawa:2020swg}. We write Eq.~\eqref{eq:Omega-swave} in a convenient form for comparison with the commonly quoted ``thermal" benchmark,
\begin{equation}
\Omega_\chi h^2 \;\simeq\; 0.12\,
\left(\frac{x_f}{25}\right)\!
\left(\frac{90}{g_\star(T_f)}\right)^{\!1/2}
\frac{2.2\times 10^{-26}\ \text{cm}^3\text{s}^{-1}}{\langle \sigma v \rangle}\,.
\label{eq:Omega_canonical_rescaled}
\end{equation}
We emphasize that the numerical choices $x_f \simeq 25$ and $g_\star \simeq 90$ commonly used in the literature are appropriate for weak scale freeze-out, but are generally not valid for sub-GeV DM~\cite{Steigman:2012nb}. In particular, when freeze-out occurs at $T_f\ll 1~\text{GeV}$, $g_\star(T_f)$ is typically ${\cal O}(10)$ rather than ${\cal O}(100)$ (e.g.\ $g_\star=10.75$ after $\mu^\pm$ annihilation and before $e^+e^-$ annihilation, dropping to $g_\star\simeq 3.36$ after $e^+e^-$ annihilation). Similarly, $x_f$ goes down from 25 to about 10 as $m_\chi$ goes from $10^4$ GeV to 100 keV. Accordingly, the $\langle\sigma v\rangle$ needed to reproduce the  observed relic density $\Omega_\chi h^2=0.120\pm 0.001$~\cite{Aghanim:2018eyx} slightly increases as we go to lower DM masses.  Therefore, we numerically solve Eq.~\eqref{eq:Omega-swave} to obtain the exact value of $|y_\chi^{i3}|$ needed for the correct relic density as a function of the DM and mediator masses. This is illustrated in Fig.~\ref{fig:money_model1} by the brown curve for a fixed ratio of the mediator and DM masses. 

\subsection{Indirect Detection} 
The direct DM annihilation to  neutrinos [cf. Eq.~\eqref{eq:424}] leads to a monochromatic neutrino spectrum at the annihilation site: 
\begin{align}
    \frac{dN_\nu}{dE}=2\frac{m_\chi}{E^2}\delta\left(1-\frac{E}{m_\chi}\right) \, .
\end{align}
The neutrino flux receives two primary contributions: a Galactic component and an extragalactic component. The Galactic contribution originates from DM annihilation into neutrinos within the Milky Way and its surrounding DM halo. This produces an anisotropic neutrino signal that is strongest toward the Galactic center. In contrast, the extragalactic component arises from neutrino emission associated with structures outside the Milky Way and includes contributions integrated over all redshifts. As a result, the corresponding neutrino signal is expected to be isotropic.

Assuming an equal flavor composition, the Galactic contribution to the neutrino flux for each flavor can be written as
\begin{align}
    \frac{d\Phi_\nu^{\rm gal.}}{dE} = \frac{1}{4\pi}\frac{\langle \sigma v\rangle}{\kappa m_\chi^2}\frac{1}{3}\frac{dN_\nu}{dE}J(\Omega) \, ,
\end{align}
where $\kappa=4~(2)$ for Dirac~(Majorana) DM, and $J\equiv \int d\Omega \int_{\rm l.o.s.}\rho_\chi^2(l)dl$ is defined as a three-dimensional integral over the observed solid angle in the sky, $d\Omega$, together with the line-of-sight (l.o.s.) distance element $dl$, involving the dark matter density $\rho_\chi(l)$ along that direction.  

The isotropic extragalactic contribution consists of two parts: one arising from the smooth, non-collapsed DM distribution, and another originating from overdense regions within galactic halos throughout the Universe. Consequently, the predicted neutrino flux can be expressed as
\begin{align}
\frac{d\Phi_\nu^{\rm exgal.}}{dE} = \frac{1}{4\pi}\frac{\rho_\chi^2\langle \sigma v\rangle}{\kappa m_\chi^2}\frac{1}{3}\int_0^{z_{\rm max}}dz\frac{[1+G(z)](1+z)^3}{H(z)}\frac{dN_\nu(E')}{dE'} \, ,
\end{align}
where $z$ is the redshift (with its maximum value $z_{\rm max}\sim 40$ beyond which the contribution to the integral becomes negligible),  $H(z)$ is the Hubble parameter and $G(z)$ is the halo boost factor at redshift $z$, and $E'_\nu$ is the neutrino energy at the detector, which is related to the neutrino energy $E_\nu$ at the source via a Jacobian transformation, namely, 
\begin{align}
    \frac{dN_\nu(E')}{dE'} = \frac{2m_\chi}{{E'}^2}\delta\left(\frac{m_\chi}{E'}-1\right) = \frac{2}{E}\delta\left[z-\left(\frac{m_\chi}{E}-1\right)\right] \, .
\end{align}
For more details, see e.g., Refs.~\cite{Arguelles:2019ouk}.

To date, no experimental observation has identified a signal associated with DM  annihilation. As a result, various experiments have placed constraints on the differential neutrino flux $d\Phi_\nu/dE$ from DM annihilation across different energy ranges, which can be translated into upper limits on the annihilation rate $\langle \sigma v\rangle$ as a function of the DM mass for a given DM density profile. Using the reported neutrino flux data from solar, atmospheric, and astrophysical neutrino observations, such upper limits on the DM annihilation cross section, $\langle \sigma v\rangle$, have been derived~\cite{Arguelles:2019ouk}, which can be translated into an exclusion region in our model parameter space using Eq.~\eqref{eq:a-Dirac}. The resulting constraint is shown in Fig.~\ref{fig:money_model1} by the green-shaded region (labeled `indirect'). The experimental limits on $\langle \sigma v\rangle$ were mostly taken from Ref.~\cite{Arguelles:2019ouk}, with recent updates from XENONnT~\cite{XENON:2022ltv, BetancourtKamenetskaia:2025rwk} and IceCube~\cite{IceCube:2025fcn} in the low-mass and high-mass regions, respectively. For related discussion on neutrino constraints on the DM annihilation, see also Refs.~\cite{Beacom:2006tt,Yuksel:2007ac,Olivares-DelCampo:2017feq, Blennow:2019fhy, ElAisati:2017ppn, Klop:2018ltd}. 

\subsection{Direct detection}
In our setup, tree–level quark–DM scattering is absent, because the Yukawa interaction in Eq.~\eqref{eq:Dirac-int} couples $\chi$ only to leptons and to the inert doublet. The only renormalizable portal between the dark sector and the SM quarks arises at one loop through penguin diagrams with SM photon, $Z$-boson, and Higgs boson~\cite{Ibarra:2015fqa,Ibarra:2016dlb,Herrero-Garcia:2018koq}. In the regime of our interest, $m_{H_2^\pm},\,m_A \ \gg\ m_{H_2^0},$ the amplitude is saturated by the $(\nu,\,H_2^0)$ loop, whereas the $(H_2^\pm,A)$ contributions becomes subdominant. Thus, the dominant elastic channel is therefore generated by the gauge–invariant sum of $Z$–penguin diagrams.   More importantly, because the dark fermion is (pseudo-)Dirac, the $Z$ exchange will also induce spin-independent vector contact interaction with quarks in addition to the spin-dependent axial contact interaction. The axial contact interaction was calculated in Ref.~\cite{Ibarra:2016dlb}, and here we translate this for the vector interaction case. The corresponding effective Lagrangian is  
\begin{equation}
\label{eq:Leff-axial}
\mathcal L_{\rm eff}\;\supset\;\zeta_q\,(\bar\chi\gamma_\mu\chi)\,(\bar q\gamma^\mu q),
\qquad
q=u,d,s,
\end{equation}
with Wilson coefficient $\zeta_q$ given by\footnote{Note that Eq. (65) of Ref.~\cite{Ibarra:2016dlb} is missing a factor of $g/2c_W$.} 
\begin{equation}
\label{eq:xiq}
\zeta_q \;\simeq\; \frac{|y_\chi^{i 3}|^2\,v_q}{128\pi^2\,m_Z^2}\,\frac{g^2}{c_W^2} c_L \,
G_2\!\left(\frac{m_\chi^2}{m_{H_2^0}^2}\right).
\end{equation}
Here $g$ is the $SU(2)_L$ gauge coupling, $c_W\equiv\cos\theta_W$, with $\theta_W$ being the weak mixing angle, and the $Z$ vector charges are $v_u = 1/2 - (4/3) s_W^2$ and $v_d = -1/2 + (2/3) s_W^2$, with $s_W^2 = \sin^2\theta_W \simeq 0.2312$~\cite{ParticleDataGroup:2024cfk}. The loop function $G_2(x)$ is given by
\begin{equation}
G_2(x)\;=\;-1+\frac{2\big[x+(1-x)\ln(1-x)\big]}{x^2}.
\end{equation}  
In the limit $x\ll 1$, $G_2(x) \simeq x/3$. We have included a logarithmic enhancement factor $c_L\sim 1+\ln(m_{A}^2/m_{H_2^0}^2)\sim {\cal O}(10)$ in Eq.~\eqref{eq:xiq} for our hierarchical scenario $m_{A}\gg m_{H_2^0}$. 

The elastic spin-independent DM direct detection cross section with a nucleus with $Z$ protons and $(A-Z)$ neutrons reads as
\begin{equation}
\label{eq:sigma-SI-A}
\sigma^{\rm SI}_A \;=\; \frac{1}{\pi} \mu_{\chi N}^2 \,\Big[\,Z\,f_V^p+(A-Z)\,f_V^n\,\Big]^2.
\end{equation}
Here $f_V^p = 2 \zeta_u + \zeta_d$, $f_V^n = \zeta_u + 2 \zeta_d$ are the form factors, and $\mu_{\chi N} = m_\chi m_N/(m_\chi+m_N)$ with $N=p,n$ is the reduced mass. We use the current leading experimental upper limits on $\sigma^{\rm SI}$ from DarkSide~\cite{DarkSide:2022dhx} (below 5 GeV), PandaX-4T~\cite{PandaX:2024qfu} (between 5 and 10 GeV), LZ~\cite{LZ:2022lsv} and XENON nT~\cite{XENON:2025vwd} (above 10 GeV) to derive an upper limit on the Yukawa coupling $|y_{\chi}^{i 3}|$ as a function of the DM mass, as shown by the orange-shaded region (labeled `direct') in Fig.~\ref{fig:money_model1}. Although we have used the full expression~\eqref{eq:sigma-SI-A} to derive this limit, it is instructive to give the approximate form in the limit $m^2_\chi\ll m^2_{H_2^0}$: 
\begin{align}
\label{eq:sigmaSI-n-num}
\sigma^{\rm SI}_n &\;\simeq\;
6.1\times 10^{-44}\ {\rm cm}^2\;
\Big(\tfrac{\mu_{\chi n}}{\rm GeV}\Big)^{\!2}
\,|y_\chi^{i 3}|^4\,
\Big(\tfrac{m_\chi}{m_{H_2^0}}\Big)^{\!4},\\[4pt]
\label{eq:sigmaSI-p-num}
\sigma^{\rm SI}_p &\;\simeq\;
3.4\times 10^{-46}\ {\rm cm}^2\;
\Big(\tfrac{\mu_{\chi p}}{\rm GeV}\Big)^{\!2}
\,|y_\chi^{i 3}|^4\,
\Big(\tfrac{m_\chi}{m_{H_2^0}}\Big)^{\!4}, \\
\sigma_A^{\rm SI}&\;\simeq\; Z^2\sigma_p^{\rm SI}+(A-Z)^2\sigma_n^{\rm SI} \, .
\label{eq:SIapprox}
\end{align}   

It is important to make a note that there is also a tree–level direct–detection channel mediated by the charged scalar $H_2^+$.  Integrating out $H_2^+$ in the $t$–channel $\chi e \to \chi e$ diagram induced by Eq.~\eqref{eq:Dirac-int} generates the effective four–fermion operator
\begin{equation}
\label{eq:Leff-tree-e}
\mathcal L_{\rm eff}^{(H_2^+)} \;\simeq\;
\frac{|y_\chi^{13}|^2}{m_{H_2^+}^2}\,
(\bar\chi_R e_L)\,(\bar e_L \chi_R)
\;\simeq\;
\frac{|y_\chi^{13}|^2}{2\,m_{H_2^+}^2}\,
(\bar\chi\gamma_\mu P_R\chi)\,(\bar e\gamma^\mu P_L e)\,,
\end{equation}
where in the second step we have used a Fierz transformation.  Treating the electron as a free target, the resulting spin–averaged elastic cross section is
\begin{equation}
\label{eq:sigmae-tree}
\sigma_e^{\rm tree}
\;\simeq\;
\frac{1}{4\pi}\,\frac{m_\chi^2 m_e^2}{(m_\chi+m_e)^2} \frac{|y_\chi^{13}|^4}{m_{H_2^+}^4}\,. 
\end{equation}
For $m_\chi\gg m_e$ one has
\begin{equation}
\sigma_e^{\rm tree}
\;\simeq\;
8.1\times10^{-44}\ {\rm cm}^2\;
|y_\chi^{13}|^4\,
\Big(\frac{100~{\rm GeV}}{m_{H_2^+}}\Big)^{\!4}.
\end{equation}
Thus, in the inert-doublet lepton-portal realization of Dirac fermion DM, the electron-mediated direct detection is  in general suppressed compared to the $Z$–penguin–induced nuclear scattering rate which is enhanced by $(A-Z)^2$ [cf.~Eq.~\eqref{eq:SIapprox}]. Moreover, if $\chi$ couples predominately to $\mu$ or $\tau$ (i.e., $y_\chi^{13} \to 0$), the tree-level electron channel is absent and the DM scattering proceeds only via the loop-induced $Z$-penguins as described earlier.  

There also exists a Higgs-mediated one-loop diagram that induces effective scalar coupling $\Lambda_q\,\bar\chi\chi\,\bar q q$. This may give sizable contribution to the direct detection cross section for  large quartic couplings $\lambda_3$ and $\lambda_4$~\cite{Ibarra:2016dlb}: 
\begin{equation}
\Lambda_q \;\simeq\; -\frac{|y^{i 3}_\chi|^2}{16\pi^2}\,\frac{m_q}{m_h^2\,m_\chi}
\left[\lambda_3\,G_1\!\Big(\frac{m_\chi^2}{m_{H_2^\pm}^2}\Big)
+\frac{\lambda_3+\lambda_4}{2} \,G_1\!\Big(\frac{m_\chi^2}{m_{H_2^0}^2}\Big)\right],
\end{equation}
where the loop factor $G_1(x)$ is given by
\begin{equation}
G_1(x)=\frac{x+(1-x)\ln(1-x)}{x}\,. 
\end{equation}
For $\lambda_{3,4}\!\lesssim\!{\cal O}(0.1)$ this is typically subdominant; for $\lambda_{3,4}\!\sim\!{\cal O}(1)$ it can reach the sensitivities of next–generation direct detection experiments.

\subsection{Model prediction}\label{sec:model1_prediction}
\begin{figure}[t!]
\centering
\includegraphics[width=0.8\textwidth]{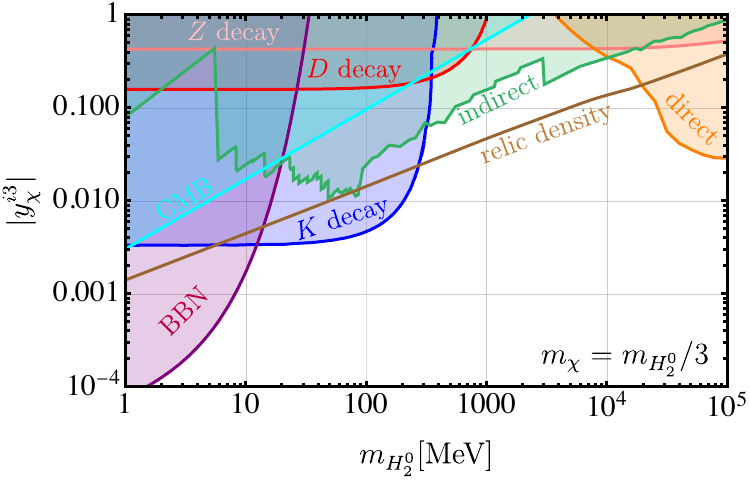}
\caption{Summary of Model~1 prediction in the mediator mass $m_{\rm med} \equiv m_{H_2^0}$ versus Yukawa coupling $y_\chi^{i3}$ plane for a benchmark mass ratio $m_\chi = m_{\rm med}/3$. The shaded regions are excluded by BBN (purple), CMB (cyan), $K$ (blue) and $D$ (red) meson decays, $Z$ invisible decay (pink), and DM direct (orange) and indirect (green) detection constraints. The  brown curve indicates the coupling values required for the thermal DM freeze-out into neutrinos that reproduces the observed relic abundance.}
\label{fig:money_model1}
\end{figure}

At energies below the heavy inert states, Dirac fermion DM $\chi$ couples predominantly to active neutrinos through the exchange of the light CP-even inert scalar $H_2^0$. The low energy phenomenology is then controlled by  3 parameters, namely, 
\begin{equation}
    m_{\rm DM}\equiv m_\chi,\qquad m_{\rm med}\equiv m_{H_2^0},\qquad y_{\chi}^{i 3},
\end{equation}
with other inert-doublet components ($A$ and $H_2^\pm$) taken to be much heavier and integrated out. In Fig.~\ref{fig:money_model1} we summarize the resulting phenomenology in the $(m_{H_2^0},\,|y_{\chi}^{i 3}|)$ plane for a representative benchmark mass ratio
$m_{\chi}=m_{H_2^0}/3$.  The brown curve labeled ``relic density'' shows the coupling required to reproduce the observed DM  abundance through $\chi\bar\chi\to \nu\bar\nu$ thermal freeze-out.\footnote{In principle, this minimal requirement   can be relaxed to open up more parameter space by going beyond the simple thermal DM picture, e.g. by invoking a freeze-in mechanism~\cite{Hall:2009bx}. Since our goal here is to find the maximum allowed neutrino-DM interaction, we do not consider such possibilities.} As shown here, for a fixed mass ratio, the required coupling increases with the mediator mass.  Similarly, the direct detection constraint discussed above is shown by the orange-shaded region labeled ``direct". It is less stringent at lower DM masses because of the reduced experimental sensitivity in this regime. We also show in Fig.~\ref{fig:money_model1} other leading constraints coming from laboratory searches for $Z$ invisible decay (pink-shaded), rare meson decays such as $K\to e\chi H_2^0$ (blue-shaded) and $D\to e\chi H_2^0$ (red-shaded), as well as the cosmological constraints of requiring both $\chi\bar\chi\to \nu\bar{\nu}$ and $\nu\bar{\nu}\to \chi\bar{\chi}$ processes to be out-of-equilibrium (i.e.~their rates must be less than the Hubble rate) at the BBN (purple-shaded), and $\chi\bar{\chi}\to \nu\bar{\nu}$ contribution to extra nonthermal neutrino energy density to be consistent with the $\Delta N_{\rm eff}$ constraint from CMB (cyan-shaded). For more details on these constraints and other sub-leading constraints on neutrino-DM interactions, see Ref.~\cite{Dev:2025tdv}. It is important to note here that the $K$ and $D$ meson decay constraints, as well as the indirect detection constraints, are flavor-dependent, and are typically less stringent for coupling to $\mu$ flavor, while they completely go away for the $\tau$ flavor. On the other hand, the $Z$ decay, BBN and CMB constraints apply to all three lepton flavors.

Fig.~\ref{fig:money_model1} shows that there is a limited range of DM mass $m_\chi\in [{\cal O}(100)~{\rm MeV},{\cal O}(10)~{\rm GeV}]$ that satisfies the correct thermal relic abundance, while being allowed by the current constraints in this model setup. Moreover, neutrino-DM couplings up to ${\cal O}(0.1)$ is allowed in this model. Future improvements in the DM direct detection cross section limits, especially in the sub-GeV DM mass range, would be crucial to probe such large neutrino-DM interactions.

Before closing this subsection, we should emphasize that the derivation of the DM-nuclear scattering rates as given above is only possible in a UV-complete model, where the mediator couplings to DM and SM are known and the loop divergences can be canceled exactly. In a DM-LEFT or DM-SMEFT version with only DM-lepton interactions, it is not possible to evaluate the loop diagrams in a consistent manner. That is why the direct detection constraints were not shown in Ref.~\cite{Dev:2025tdv}.

\section{Other DM models}
\label{sec:5}
Before discussing other viable models with potentially large neutrino-DM interactions, it is worth emphasizing that not every UV completion of the DM-SMEFT operators yields the desired low-energy $\chi\chi\nu \nu$ interactions. For instance, consider a model for Majorana singlet DM $\chi$ that contains a vector-like charged singlet fermion $\psi_2 (1,1,0)$ and a complex charged singlet scalar $\phi_2(1,1,1)$. A stabilizing $\mathbb{Z}_2$-parity can be imposed with $\{\chi,\phi_2^+\}\to-\{\chi,\phi_2^+\}$ and all SM fields and $\psi_2$ are taken to be $\mathbb{Z}_2$–even. This assignment forbids $LH\chi$ and $\phi_2 H^\dagger H$ terms. Integrating out the $\{\phi_2, \psi_2\}$ fields will generate the Majorana DM-SMEFT operator ${\widetilde{O}}_{8M}^3$ via the topology ${\cal T}_8^8$. However, this setup does not lead to the desired DM-LEFT operator $\nu_L \chi_L \nu_{\ R}^c \chi_{\ R}^c$, instead it induces operator with charged leptons, enabling $e^+ e^- \to \chi \chi$ interaction. Such effective interaction is constrained by LEP mono-photon searches~\cite{L3:1998uub, L3:2003yon}), as well as enhanced $\chi-e$ scattering at tree level, constrained by DM-electron scattering experiments~\cite{SENSEI:2023zdf, DAMIC-M:2025luv}. 

In the following subsections, we discuss two concrete models for Majorana fermion DM that can generate potentially large neutrino-DM interactions. The case for Dirac DM is effectively realized within the scotogenic framework discussed earlier, where DM candidate is a pseudo-Dirac fermion. We also briefly discuss the UV-completion of a scalar DM model and comment on the possibility of vector DM.  These discussions are intentionally kept brief and are included primarily for completeness. A  comprehensive analysis of these models, as well as the UV completion of other effective operators listed here, is left for future work. 

\subsection{Model 2: 
 Majorana DM with type-II seesaw}

In this setup we extend the SM by a real singlet scalar $\phi_1 (1,1,0)\equiv S $ and a complex hypercharge-$Y=1$ triplet $\phi_5 (1,3,1)\equiv \Delta$. We impose a stabilizing $\mathbb{Z}_2$ under which $\chi\to-\chi$ while $S$, $\Delta$, and all SM fields are $\mathbb{Z}_2$-even. This forbids renormalizable operators with a single $\chi$ insertion (such as $L\!\cdot\!H\,\chi$), ensuring the stability of $\chi$, while allowing the renormalizable interactions relevant for tree-level matching. The corresponding interaction Lagrangian is given by 
\begin{align}
\mathcal{L}_{\text{int}}
\;\supset\;&
\Big[\frac{1}{2}\,y_S\, S\,\chi^T C\,\chi
\;+\;
\frac{1}{2}\,y_\Delta^{ij}\, L_i^T C\, i\sigma_2\,\Delta\, L_j
\;+\;
\kappa\, S\, H^T i\sigma_2\,\Delta^\dagger H
\;+\;
\mu_\Delta\,H^T i\sigma_2 \Delta^\dagger H
\nonumber\\
&- \frac{1}{2} m_\chi\ \chi^T C \chi \;+\;\text{H.c.}\Big] + \mu_S\, S\, \text{Tr}(\Delta^\dagger \Delta)
\;-\;
\frac{1}{2}M_S^2\,S^2
\;-\;
M_\Delta^2\,\text{Tr}(\Delta^\dagger\Delta) \,,
\label{eq:L_model2}
\end{align}
where $y_\Delta^{ij}=y_\Delta^{ji}$ is symmetric in flavor. Additional quartic terms in the scalar potential are present but are not needed for the tree-level matching discussed below.

\begin{figure}
    \centering    \includegraphics[width=0.2\linewidth]{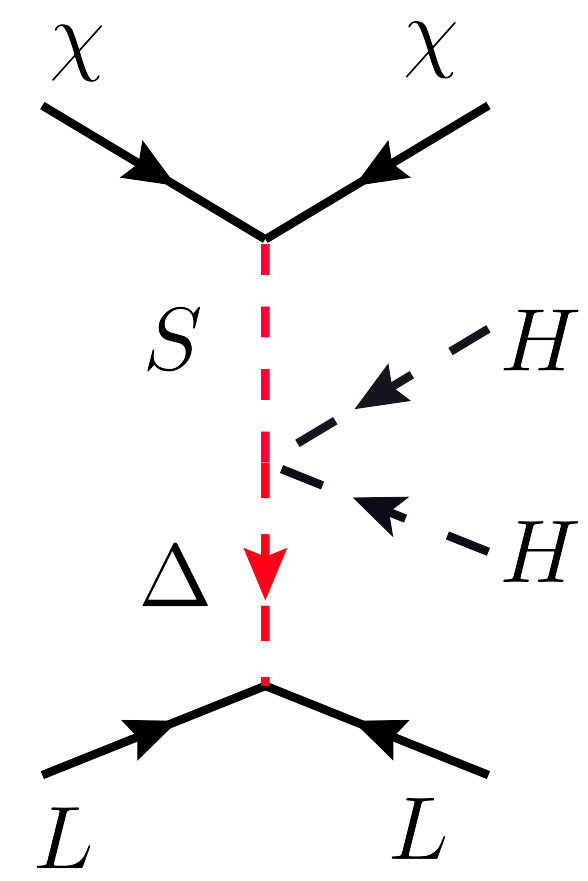}
    \hspace{20mm}\includegraphics[width=0.3\linewidth]{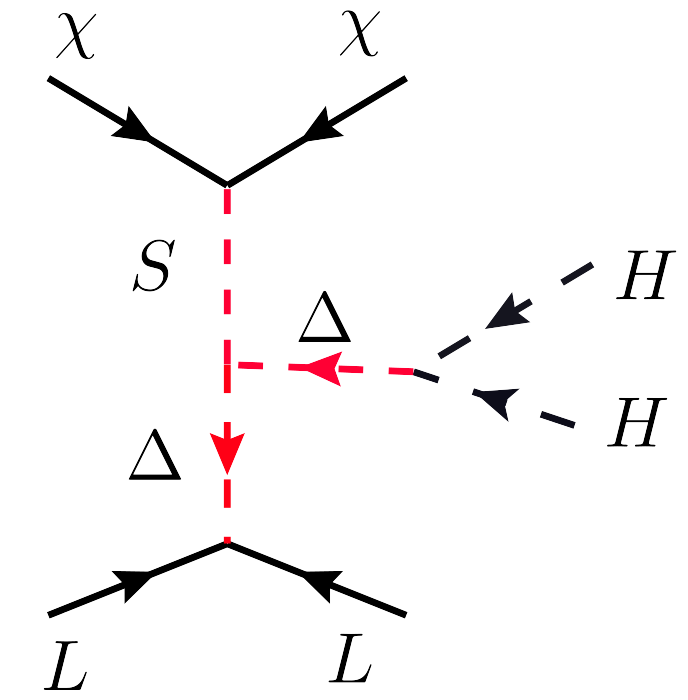}
    \caption{Feynman diagrams illustrating the UV completions of effective operators ${\widetilde{\cal O}}_{8,M}^1 \equiv (\vec L\cdot \vec H\ \vec L\cdot \vec H) \chi_L \chi_L$ and ${\widetilde{\cal O}}_{8,M}^2 \equiv (\vec L\cdot \vec H\ \vec L\cdot \vec H) \chi^c_{~R} \chi^c_{~R}$ arising in Model~2, which introduces an additional singlet scalar $S$ and a triplet scalar $\Delta$. }
    \label{fig:model2}
\end{figure}

Integrating out $S$ and $\Delta$ at tree level generates the dimension-8 DM-SMEFT operators $\widetilde{\mathcal O}^{\,1}_{8,M}$ and $\widetilde{\mathcal O}^{\,2}_{8,M}$ of Table~\ref{tab:operators}. There are two representative ways to connect the external fields $(LL)(HH)(\chi\chi)$, as shown in Fig.~\ref{fig:model2}:
\begin{itemize}
\item[(i)] \textbf{Single-$S$/single-$\Delta$ exchange} (topology ${\cal T}_8^6$), where one
$\Delta$ propagator links the $y_\Delta\,LL\Delta$ and $\kappa\,S\,HH\Delta^\dagger$ vertices,
and the singlet $S$ then links to $\chi\chi$ via $y_S$.

\item[(ii)] \textbf{Two-$\Delta$ chain} (topology ${\cal T}_8^1$), where the singlet couples to a
$\Delta^\dagger\Delta$ pair via $\mu_S S\,\mathrm{Tr}(\Delta^\dagger\Delta)$; one $\Delta$
propagator ends on $LL$ through $y_\Delta$, while the other ends on $HH$ through the
trilinear $\mu_\Delta HH\Delta^\dagger$. This contribution is therefore suppressed by an
additional $1/M_\Delta^2$ relative to (i).
\end{itemize}

With these interactions, the resulting Wilson coefficients can be written as
\begin{equation}
\label{eq:C81_model2}
\big(\widetilde C^{\,1}_{8,M}\big)^{ij} \;=\; \big(\widetilde C^{\,2}_{8,M}\big)^{ij}
\;\simeq\;
\frac{y_S\,y_\Delta^{ij}}{4}\left[
\frac{\kappa}{M_S^{2}\,M_\Delta^{2}}
\;+\;
\frac{\mu_\Delta\, \mu_S}{M_S^{2}\,M_\Delta^{4}}
\right],
\end{equation}
The equality $\widetilde C_{8,M}^1=\widetilde C_{8,M}^2$ reflects the fact that the Majorana bilinear $\chi^T C\chi$ contains both chiral structures that map onto the two SMEFT operators in
Table~\ref{tab:operators}. Here the first term is the single–$S$–single–$\Delta$ exchange (${\cal T}_8^6$), while the second term arises from the $\mu_S$–$\mu_\Delta$ chain with two $\Delta$ propagators (${\cal T}_8^1$), hence the extra $1/M_\Delta^{2}$ suppression discussed above. After EWSB, $H\to(0,(v+h)/\sqrt2)^T$, 
\begin{equation}
\widetilde{\mathcal{O}}^{\,1}_{8,M}\;\to\; \frac{v^2}{2}\,(\nu_{Li}^T C\nu_{Lj})\,(\chi^T C\chi)\,,
\qquad
(C^{\,1}_{6,M})^{ij} \;=\; \frac{v^2}{2}\,(\widetilde C^{\,1}_{8,M})^{ij}\,,
\label{eq:model2_match_LEFT}
\end{equation}
and analogously for $\widetilde{\mathcal{O}}^{\,2}_{8,M}$.

\paragraph{Relation to type-II seesaw:}
The same triplet $\Delta$ also generates neutrino masses via the usual type-II seesaw relation~\cite{Magg:1980ut,Schechter:1980gr,Lazarides:1980nt,Mohapatra:1980yp}. The trilinear $\mu_\Delta HH\Delta^\dagger$ induces a triplet VEV
\begin{equation}
v_\Delta \ \simeq\ \frac{\mu_\Delta\,v^2}{2 M_\Delta^2}\,,
\label{eq:vdel}
\end{equation}
and the neutrino mass matrix reads as
\begin{equation}
(m_\nu)_{ij}\ = y_\Delta^{ij}\,v_\Delta\,.
\label{eq:nutrip}
\end{equation}
Electroweak precision constraints on the $\rho$ parameter require $v_\Delta\ll v$~\cite{ParticleDataGroup:2024cfk}, and for a hypercharge $Y=1$ triplet typically imply $v_\Delta\lesssim \mathcal{O}(\text{GeV})$~\cite{Kanemura:2012rs}. Using Eqs.~\eqref{eq:vdel} and \eqref{eq:nutrip}, one may rewrite
Eq.~\eqref{eq:model2_match_LEFT} as
\begin{equation}
(C_{6,M}^{1})^{ij}\ \simeq\
\frac{y_S}{4M_S^2}
\left[
\frac{\kappa}{\mu_\Delta}
+\frac{\mu_S}{M_\Delta^2}
\right](m_\nu)_{ij}\,,
\label{eq:C6_in_terms_mnu}
\end{equation}
which makes explicit that the effective $(\nu\nu)(\chi\chi)$ interaction is tied to the same flavor structure that controls $(m_\nu)_{ij}$.

\paragraph{Two useful regimes:}
The phenomenology depends strongly on whether neutrino masses are realized with small $y_\Delta$ and sizable $v_\Delta$, or with $y_\Delta\sim \mathcal O(1)$ and a tiny induced VEV $v_\Delta$.

\begin{itemize}
\item \textbf{$y_\Delta \ll 1$ with $v_\Delta \sim {\cal O \text{(GeV)}}$:}
Taking $v_\Delta$ near its electroweak precision upper limit implies $y_\Delta^{ij}\sim (m_\nu)_{ij}/v_\Delta \ll 1$, which automatically suppresses charged-lepton flavor violation. In this regime the triplet decays are typically dominated by gauge-boson modes (e.g.\ $\Delta^{\pm\pm}\to W^\pm W^\pm$ and cascade modes (depending on the mass splitting $\Delta m=m_{H^{\pm\pm}}-m_{H^\pm}$ between doubly- and singly-charged components)~\cite{FileviezPerez:2008jbu, Melfo:2011nx}) rather than dileptons, and collider bounds on $\Delta^{\pm\pm}$ can be substantially weaker than in the leptonic-decay regime. In fact, there is currently a mass gap between $(84-200)$ GeV [or $(55.6-200)$ GeV for $\Delta m<0$] for doubly charged scalars~\cite{Kanemura:2014goa, 
Ashanujjaman:2022ofg, Ashanujjaman:2025scr}, whereas the mass range $(200-400)$ GeV is excluded~\cite{ATLAS:2021jol,Ashanujjaman:2021txz}. The neutral scalar $\Delta^0$ can be even lighter which can be understood from the following mass relation~\cite{Mandal:2022zmy} (in the limit of $v_\Delta \ll v$)
\begin{equation}
    m_{\rm Re(\Delta^{0})}^2 \simeq m_{\rm Im(\Delta^{0})}^2 = m_{\Delta^+}^2 + \frac{\lambda}{4} v^2 = m_{\Delta^{++}}^2 + \frac{\lambda}{2} v^2 \, ,
\end{equation}
where $\lambda$ arises from the quartic coupling $H^\dagger \Delta \Delta^\dagger H$. However, if the CP-odd and CP-even states are nearly degenerate, there is a strong limit from $Z$-boson decay $Z_\mu \to \Delta^0 \Delta^{0*}$, $m_{\Delta^0} > 45.6$ GeV~\cite{LEP:2000pgt}. 

For the neutrino-DM interaction, however, Eq.~\eqref{eq:C81_model2} and Eq.~\eqref{eq:C6_in_terms_mnu}  shows that the LEFT coefficient remains parametrically small, essentially because the same small $y_\Delta$ that suppresses cLFV also suppresses the $(\nu\nu)(\chi\chi)$ operator. In particular, even if the singlet field $S$ is taken to be light, the effective coupling inherits the $m_\nu$ suppression through $y_\Delta$.

\item \textbf{$y_\Delta\sim \mathcal{O}(1)$ with $v_\Delta\ll 1~\text{GeV}$:}
Alternatively, one may take $y_\Delta$ to be order unity and generate small neutrino masses with a very small induced VEV $v_\Delta\sim (m_\nu/y_\Delta)\ll 1~\text{GeV}$, which corresponds to a very small $\mu_\Delta$ (technically natural as lepton number is restored in the limit $\mu_\Delta\to 0$). In this regime the doubly-charged scalar typically decays dominantly into same-sign dileptons $\Delta^{\pm\pm}\to \ell^\pm_i \ell^\pm_j$, leading to strong collider constraints excluding masses below 1020 GeV for $e/\mu$-dominated textures~\cite{ATLAS:2014kca,ATLAS:2017xqs, ATLAS:2022pbd}, while $\tau$-aligned textures remain somewhat less constrained with a lower bound of 535 GeV~\cite{CMS:2017pet}.

Even though the triplet field $\Delta$ is at ${\cal O}({\rm TeV})$ scale, the scalar singlet field $S$ can be taken arbitrarily light.  A particularly transparent limit is $m_\chi \ll m_S \ll v \ll M_\Delta$,
in which the triplet can be integrated out first (in the unbroken theory), generating the effective
dimension-4 coupling originating from an SMEFT dimension-6 term:
\begin{equation}
\mathcal L_{\rm eff}\ \supset\
\frac{1}{2}\,g_{S\nu\nu}^{ij}\,S\,\nu_{Li}^T C\nu_{Lj} + \text{H.c.},
\qquad
g_{S\nu\nu}^{ij}\ \simeq\
\frac{y_\Delta^{ij}}{2}\left(\frac{\kappa v^2}{M_\Delta^2}+\frac{\mu_\Delta\mu_S v^2}{M_\Delta^4}\right).
\label{eq:gSnunu_model2}
\end{equation}
Here one can ignore the second term in $g_{S\nu\nu}$ as $\mu_\Delta \ll M_\Delta$. Integrating out $S$ at $\mu\simeq m_S$ then yields the DM-LEFT contact interaction
\begin{equation}
(C_{6,M}^{1})^{ij}\ \simeq\ \frac{y_S\,y_\Delta^{ij} \kappa v^2}{8 m_S^2 M_\Delta^2}\,,
\label{eq:LLEFT_model2_lightS}
\end{equation}
which makes explicit the enhancement with a lighter singlet mediator.  For comparison, setting $m_{S} = 50$ MeV and Yukawa couplings $y_S =1, y_\Delta = 0.1$, $\kappa =1$, $m_\Delta = 1$ TeV leads to a sizable enhancement of neutrino-DM interaction, corresponding to $C_{6,M}^{1} \simeq 2.6\times 10^{4}\ G_F$.  

\end{itemize}

\subsection{Model 3: Majorana DM with inverse seesaw}

This model introduces a Majorana fermion DM $\chi$, as well as a complex scalar singlet $\phi_1(1,1,0)$ and a vector-like fermion $\psi_1(1,1,0)$.  This setup has a natural embedding in the inverse seesaw model of neutrino masses~\cite{Mohapatra:1986bd}, with the $\psi_1(1,1,0)$ identified as the $\nu^c$ and $N$ fields.  Therefore, we elaborate on the extended inverse seesaw model here, which contains three families of $\nu^c_i$ and $N_i$ fields. The extension is in the addition of the Majorana DM fermion $\chi(1,1,0)$ as well as the complex scalar singlet $\phi_1(1,1,0) \equiv S$, compared to the conventional inverse seesaw model. A softly broken lepton number symmetry is assumed, as well as a $\mathbb{Z}_2$ symmetry that stabilizes the DM.  The quantum numbers of the lepton fields and other SM-singlet fields under $U(1)_L \times \mathbb{Z}_2$ are shown in Table~\ref{tab:tab7}. 

\begin{table}[h!] 
   \small
   \centering
   \begin{tabular}{|c|c|c|c|c|c|}
   \hline\hline
   \textbf{Field} & $L_i(1,2,-\frac{1}{2})$ & $\nu^c_i(1,1,0)$ & $N_i(1,1,0)$ & $\chi(1,1,0)$ & $S(1,1,0)$ \\ 
   \hline
  $U(1)_L$ charge &  $+1$ & $-1$ & $+1$ & $+1$ & $+2$ \\ \hline
  $\mathbb{Z}_2$ charge & $+1$ & $+1$ & $+1$ & $-1$ & $+1$  \\
   \hline\hline 
   \end{tabular}
   \caption{Global lepton number charges $U(1)_L$ of leptons and new SM-singlet particles introduced in Model 3. Also shown are their $\mathbb{Z}_2$ charges. All other SM fields are neutral under $U(1)_L$ as well as $\mathbb{Z}_2$. }
   \label{tab:tab7}
\end{table}

\begin{figure}[h!]
    \centering
    \includegraphics[width=0.3\linewidth]{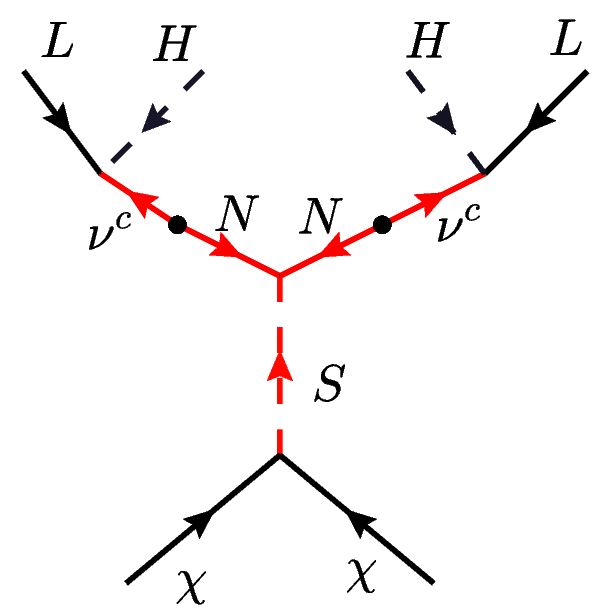}
    \caption{Feynman diagram illustrating the UV completion of the effective operators ${\widetilde{\cal O}}_{8,M}^1 \equiv (\vec L\cdot \vec H\ \vec L\cdot \vec H) \chi_L \chi_L$ and ${\widetilde{\cal O}}_{8,M}^2 \equiv (\vec L\cdot \vec H\ \vec L\cdot \vec H) \chi^c_{~R} \chi^c_{~R}$ arising in Model 3, which introduces fermion singlet fields $\{ \nu^c, N\}$ and a complex scalar field $S$. }
    \label{fig:model3}
\end{figure}

The Yukawa Lagrangian invariant under the $U(1)_L \times \mathbb{Z}_2$ relevant for neutrino and DM interactions is given by
\begin{equation}
-{\cal L}_{\rm Yuk}  \supset Y_\nu L \nu^c H + M_N \nu^c N + \frac{Y_N}{2} N N S^* + \frac{Y_\chi}{2} \chi \chi S^* + {\rm H.c.}
\end{equation}
With the choice of a negative sign for the coefficient of the $|S|^2$ term in the Higgs potential, 
$S$ will acquire a nonzero VEV, $\langle S \rangle = v_S$ and lepton number will be spontaneously broken. 
We assume that lepton number is also softly broken by a mass term $\mu^2 S^2$ in the scalar potential.  
The would-be Majoron can have an arbitrarily small mass in this case. 
The $3 \times 3$ block matrix for the neutral lepton masses spanning $(\nu, \nu^c, N)$ fields is given by
\begin{eqnarray}
M_\nu = \left(\begin{matrix}  0 & Y_\nu v/\sqrt{2} & 0 \cr Y_\nu^T v/\sqrt{2} & 0 & M_N \cr 0 & M_N^T & Y_N v_S \end{matrix}   \right)~.
\end{eqnarray}
The effective $d=6$ operators of the type $(\nu \nu)  (\chi \chi)$ are induced through the diagram shown in Fig.~\ref{fig:model3}.  The coefficient of these operators is given by
\begin{equation}
C_{6,M}^2 = \frac{(Y_\nu)^2 v^2 Y_N Y_\chi}{16 M_N^2}\left(\frac{1}{M_{S_R}^2} +\frac{1}{M_{S_I}^2}\right) \simeq \frac{\theta_{\nu N}^2 Y_N Y_\chi}{8}\left(\frac{1}{M_{S_R}^2} +\frac{1}{M_{S_I}^2}\right) .
\end{equation}
Here, $\theta_{\nu N} \simeq Y_\nu v/(\sqrt{2}M_N)$ is the $\nu-N$ mixing angle, $M_{S_R}$ and $M_{S_I}$ stand for the masses of the real and imaginary components of the complex scalar field $S$.

There are a few constraints to consider. Firstly, $Y_N v_S \sim$ a few eV is needed to reproduce the right order-of-magnitude for the neutrino mass, assuming that $Y_\nu v  \lesssim M_N$. Secondly,  $\theta_{\nu N}$ should satisfy the experimental constraints of a few to 10 percent, depending on the neutrino flavor~\cite{Bolton:2019pcu}.  And thirdly, the DM mass is given by $m_\chi = Y_\chi v_S$. A benchmark point that satisfies these constraints is given as follows:  Take $\theta_{\nu N} = 0.1, \, v_S = 10$ MeV, $M_\chi = 10$ MeV,  $Y_N = 10^{-6}, M_{S_R} = M_{S_I} = 30$ MeV. The effective coefficient with this choice is close to $G_F$, without much restrictions arising from the electron coupling. As in the case of scotogenic model, this model also will induce $\chi \chi e^+ e^-$ vertex through loops, originating from the $\nu-N$ mixing.  However, these couplings are very weak, as they involve the Yukawa coupling $Y_N \sim 10^{-6}$ required by the neutrino mass. Note that the DM field $\chi$ will decouple from the SM sector in the  $Y_N \rightarrow 0$ limit.

\subsection{Model 4: Scalar DM}
\label{sec:scalardm}
\begin{figure}
    \centering
    \includegraphics[width=0.15\linewidth]{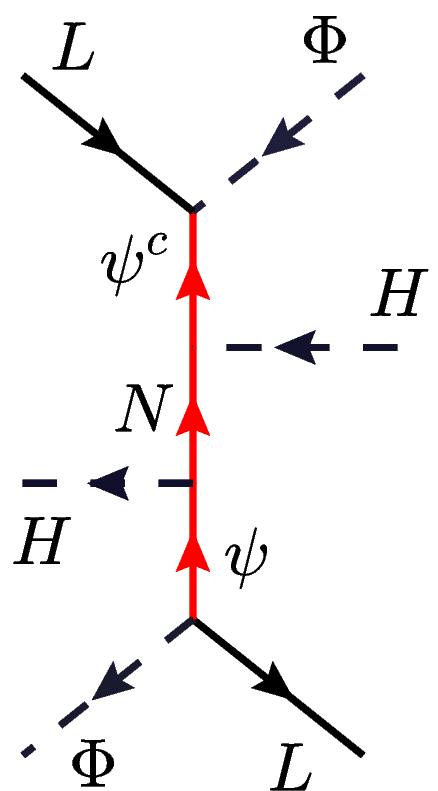}
    \caption{Feynman diagram illustrating the UV completion of the effective operator ${\cal \widetilde{O}}_{8,C}^1 \equiv (H^\dagger \bar L \gamma_\mu \vec L \cdot \vec H)\ (\partial^\mu \Phi^\dagger) \Phi$ arising in Model 4, which introduces vector-like lepton doublet $\psi_3\equiv\psi$ and a chiral fermion $\psi_1 \equiv N$.}
    \label{fig:model4}
\end{figure}
For completeness, we now discuss a model with scalar DM. Here we introduce a vector-like lepton doublet $\psi_3 (1,2,1/2)\equiv \psi$ and a Weyl fermion singlet $\psi_1 (1,1,0)\equiv N$, denoted as
\begin{eqnarray}
\psi(1,2,-\frac{1}{2}) = \left(\begin{matrix} \nu_E \cr E \end{matrix}  \right),~~~\psi^c(1,2,\frac{1}{2})= \left(   \begin{matrix}-E^c \cr \nu_E^c   \end{matrix} \right),~~~N(1,1,0)\, ,
\end{eqnarray}
with all fields being left-handed. In addition, a complex scalar field $\Phi(1,1,0)$ which carries lepton number of $-1$ is also introduced. The Lagrangian relevant to the neutrino-DM interaction is given by
\begin{eqnarray}
-{\cal L}_{\rm Yuk} =&& y_i L_i^\alpha \psi^{c \beta} \Phi \epsilon_{\alpha \beta} + M_\psi \psi^\alpha \psi^{c \beta} \epsilon_{\alpha \beta} + \lambda_1 \psi_i^\alpha H^\beta N \epsilon_{\alpha \beta} - \lambda_2 \psi^{c \alpha} \widetilde{H}^\beta N \epsilon_{\alpha \beta}\nonumber\\
&& + \frac{M_N}{2}N N + {\rm H.c.} 
\end{eqnarray}
Here $\alpha,\beta$ are $SU(2)_L$ indices.  Note that lepton number is conserved by this Lagrangian, as can be seen by assigning lepton numbers of $+1$ for $L_i$, $-1$ for $\Phi$ and zero for $(\psi, \,\psi^c,\, N)$ fields. This Lagrangian also preserves a discrete $\mathbb{Z}_2$ symmetry under which $(\Phi,\, \psi, \,\psi^c,\, N)$ are odd, while all SM particles are even. This guarantees the stability of DM $\Phi$, which is assumed be the lightest of all $\mathbb{Z}_2$-odd states. This model resembles the doublet-singlet fermionic DM model studied extensively in the literature (see e.g., Ref.~\cite{Cohen:2011ec}), but with the distinction that the singlet-doublet fermion sector serves as the mediator and not as the DM in the present case.

The mass matrix for the neutral leptons can be written down in the basis $(\nu_E, \nu_E^c, N)$ as
\begin{eqnarray}
 M_{\rm neutral} = \left(\begin{matrix} 0 & M_\psi & \frac{\lambda_1 v}{\sqrt{2}}  \cr M_\psi & 0 &  \frac{\lambda_2 v}{\sqrt{2}}   \cr   \frac{\lambda_1 v}{\sqrt{2}}   &   \frac{\lambda_2 v}{\sqrt{2}}  &  M_N\end{matrix}   \right)   
 \label{eq:21DM}
\end{eqnarray}
In contrast, the charged lepton $E$ is unmixed and has a mass given by $M_E = M_\psi$. Direct searches for vector-like charged leptons at the LHC set limits of order 700 GeV on the mass of $E$. In order to obtain large $\nu \nu \Phi \Phi$ interactions, we choose the mass of the singlet fermion $N$ to be much smaller than the EWSB scale.  To get simple analytic expressions, let us work in the limit where $\lambda_1 \rightarrow 0$ with the hierarchy $M_N \ll \lambda_2 v/\sqrt{2} \ll M_\psi$.  This will result in $\nu_E-N$ mixing given by the angle
\begin{equation}
\theta_{\nu_EN} \simeq \frac{\lambda_2 v}{\sqrt{2} M_\psi}~.
\end{equation}
The masses of $N$ and the heavier fields $(\nu_E, \,\nu_E^c)$ are given by
\begin{equation}
M_N \simeq M_N,~~M_{\nu_E} \simeq M_{\nu_E^c} = \sqrt{M^2_\psi + \lambda_2^2 v^2/2} \simeq M_\psi(1+\frac{\lambda_2^2 v^2}{4 M_\psi^2}) \simeq M_\psi(1+\frac{\theta_{\nu_EN}^2}{2}).
\end{equation}
Now, the oblique electroweak precision parameter $T$ requires the mass-splittings between the two members of a leptonic doublet to be $(m_2-m_1)^2 \leq (72~{\rm GeV})^2$, which translates into a limit of $\theta_{\nu_EN}^2 \leq 1/5$, for $M_\psi = 720$ GeV.  We shall impose this condition in estimating the $\nu\nu\Phi\Phi$ interaction strength.

The diagram shown in Fig.~\ref{fig:model4} will induce large $\nu\nu\Phi\Phi$ interactions of the form
\begin{equation}
\overline{\nu}_{iL} \gamma_\mu \nu_{jL} (\partial^\mu \Phi) \Phi^*
\end{equation}
with an amplitude $C$ given by
\begin{equation}
C_{ij} =\frac{\theta_{\nu_EN}^2 y_i y_j^*}{M_N^2}~.
\end{equation}
Since the Yukawa couplings $y_i$ are only loosely constrained, and since the singlet fermion mass $M_N$ can be as low as a 100 MeV or so, the effective $\nu\nu\Phi\Phi$ coupling can be quite large. Note that the effective $ee\Phi\Phi$ coupling is suppressed by a factor of $(M_N^2/M_E^2) \ll 1$.

\subsection{Comment on Vector DM}
Finally, the case of vector DM is more challenging to UV-complete because the mediator must be charged under a gauge symmetry in order to couple to a vector field. A consistent UV-completion requires an underlying gauge symmetry and symmetry-breaking sector, which introduces additional theoretical structure and phenomenological constraints. In fact, as recently shown in Ref.~\cite{Shaikh:2026ixs}, UV-complete realizations of vector DM can qualitatively change the phenomenological conclusions derived from an EFT description. 
We leave a detailed exploration of vector DM models for future work.  

\section{Conclusions}
\label{sec:6}
We have presented a general framework to systematically study neutrino-DM  interactions, both in the EFT approach and in UV-complete models.  Starting from a low-energy DM-LEFT description and its gauge-invariant embedding in DM-SMEFT, we identified the leading non-derivative operator structures that couple neutrinos to fermionic, scalar, and vector DM candidates. We then classified the renormalizable tree-level UV completions of the relevant DM-SMEFT operators up to dimension-8 by enumerating the distinct diagram topologies and corresponding mediator particle quantum numbers, thereby providing a clear connection between the effective interactions and minimal UV field-content. 

Using this EFT to UV framework, we constructed a few benchmark model realizations that can yield sizable neutrino-DM interactions, while avoiding potentially problematic charged-lepton interactions at the leading order, as well as the neutrino mass constraints. In particular, the inert-doublet scotogenic  realization of a pseudo-Dirac DM with a light CP-even neutral scalar mediator  illustrates how neutrino-dominated DM phenomenology can arise in this model and how a combination of the existing cosmology, flavor, electroweak precision, collider, and direct detection constraints carve out viable regions of parameter space with clear targets for future searches. In this model, we find  that an effective neutrino-DM coupling as large as $10^3 G_F$ is allowed.  

We also provided alternative model realizations for Majorana DM based on the type-II seesaw and inverse seesaw frameworks. The type-II version can give sizable neutrino-DM effective interactions as large as $10^5 G_F$, while the inverse-seesaw version can give effective couplings comparable to $G_F$. For completeness, we also discussed a simple model with scalar DM, where the effective neutrino-DM coupling can be much larger than $G_F$.  Given the large number of possible UV-completions of the effective operators listed here, we hope the model-building community can follow up with other interesting model realizations for large neutrino-DM interactions, thus shedding more light on the plausible  connection between two of the least understood  sectors in fundamental physics.    

\acknowledgments
We wish to acknowledge the Center for Theoretical Underground Physics and Related Areas (CETUP*) and the
Institute for Underground Science at SURF for hospitality and for providing a stimulating environment during the 2025 Summer Workshop (Neutrino Weeks), where
this work was initiated. K.S.B. and B.D. also thank the organizers of WHEPP 2025 at IIT, Hyderabad for local
hospitality, where a part of this work was done. B.D. thanks Doojin Kim, Deepak Sathyan, Kuver Sinha and Yongchao Zhang for  discussions on neutrino-DM interactions. A.T. thanks Julian Heeck for discussions on various topologies. The work of K.S.B. is supported by the US Department of Energy under grant
number DE-SC0016013. The work of B.D. was partly supported by the US Department of Energy under grant No.
DE-SC0017987 and by a Humboldt Fellowship from the
Alexander von Humboldt Foundation. The work of A.T. is supported in part by
the US Department of Energy grant DE-SC0010143.   

\bibliographystyle{JHEP}
\bibliography{main}
\end{document}